# Community of Practice: A framework for understanding identity development within informal physics programs

Claudia Fracchiolla[1], Brean Prefontaine[2], and Kathleen Hinko[2]
[1]University College Dublin, Belfield, Dublin 4, Ireland.
[2]Michigan State University, East Lansing, Michigan 48824, USA.

Studies on physics identity have shown that it is one of the main factors that can predict a person's persistence in the field; therefore, studying physics identity is critical to increase diversity within the field of physics and to understand what changes can allow more women and minorities to identify with the field. In this study, we investigate informal physics programs as spaces for physics identity exploration. These programs provide unique conditions under which to study physics identity development along with other identities. Informal physics spaces allow for voluntary engagement, as well as elements of agency and autonomy within the exploration of physics. Thus these spaces allow an identity to form outside of the constraints traditionally found in academic settings. In this work, we operationalized the Community of Practice (CoP) framework to study the development of physics identities within university students who facilitate informal physics programs. We present the stories from two physics graduate students out of our sample to provide a context for testing the feasibility of the extended framework and to identify how experiences within an informal physics program can shape physics identity development. This paper presents the operationalized constructs within the Community of Practice framework, how these constructs are applied to the narrated experiences of our participants, and highlights how we can use this framework to understand the nuances of physics identity development as well as the factors that can influence that development.

## Introduction

The physics field has been grappling with issues of diversity, equity, and inclusion for some time. Statistically, women and people of color are significantly underrepresented in the field compared to the US and college populations [1-2]. Members of these groups are significantly more likely to encounter external environmental conditions of harassment, bias, and hostile climates [3-8] along with internal struggles of stereotype threat, imposter syndrome, a lack of a sense of belonging, and issues around fixed mindset [9-12]. However, some initiatives have been put in place to address these issues. In particular, physics organizations around the world (IOP, EPS, AAPT, and APS) have been developing strategies to recruit more women and members of underrepresented minorities including the APS Bridge Program, the Athena Swan Charter, women in physics conference series (ICWIP, CUWiP), and organizations such as APS and AAPT adopting codes of conduct at professional meetings [13-14]. Locally, however, physics departments must make efforts to support students, staff, and faculty who may be marginalized and oppressed [15-16].

Identity, and more specifically a person's self-association with physics, has been identified as leading factors in predicting a person's future career path involving physics [17-20]. Factors such as student attitude, self-efficacy, agency, a sense of belonging, and motivation are understood to be important for building identity and addressing the number and diversity of students who persist in STEM subjects [9,21-23]. However, for students of color, women, and other marginalized groups coming into the field, building a physics identity necessarily intersects with their other identities and experiences [3, 24-26]. In some cases these intersecting identities may not align with current cultural norms and perceptions established by some members of the field surrounding who should be a physicist [26-28]. For example, a study conducted by Hyater-Adams et al. used the Critical Physics Identity framework (CPI) to look at how the experiences and structures impact the formation of black physicists identity. They found that both internal thoughts around who can be a physicist and external ideals put forth by the community and others external to the community about who can engage in physics discouraged participation in physics and therefore negatively impacting physics identity [25-26]. Thus, a systemic lack of support for physics undergraduate and graduate students' development of identity and sense of belonging in physics (so they can see themselves as a physicist) negatively impacts their career path, this is particularly true for those from underrepresented groups in physics [4-6,9,12,24,26,29]. If internal and external recognition of having the ability to be a physicist has the biggest impact on physics identity development, then it is clear that the current climate of physics allows members of marginalized groups to struggle to participate and persist in physics, preventing changes in the landscape of the field. Without the consideration of intersecting identities, these efforts will not be successful in supporting the members of physics who experience marginalization and oppression.

In our previous work, we have looked at the self-reported motivation of university students who volunteer in physics outreach[1] programs [39] and have characterized the teaching practices of volunteers in an afterschool program for youth [40]. The results of that work have motivated us to consider what it means for someone to have an experience volunteering in an informal physics program and to look at how that experience might affect their physics identity. In this work we propose to study identity development in the context of informal learning spaces through a Community of Practice (CoP) framework. Informal learning (sometimes referred to as "free-choice" learning) is characterized by learner agency and is often learner-centered. This means that participants can opt in or out of the experiences and that they often have a say in the content and how they can engage with the content. Informal physics learning can take place in a variety of settings and can include events or activities that physicists might call "outreach" or "public engagement." Many common physics outreach programs involve two groups of participants: "audiences" or members of the public and "facilitators" who are usually physicists and physics students.

While a large body of research in science education is focused on the impact that participation has on the youth and public audiences in informal learning environments, in this work we postulate to look at how volunteering to facilitate these programs impacts the identity of facilitators. We hypothesize that the environments promoted by informal learning programs allows facilitators to explore the intersection of their different identities, including their physics identity. A form of understanding and supporting intersectional identity development is through membership in Communities of Practice (CoPs) [16,30]. Therefore, in this paper we propose that informal physics programs can operate as CoPs and participation in these programs can provide positive experiences resulting in physics identity growth for the university physics student facilitators. These experiences might include the following: connections with peers and other members of the physics community, expansion of social and professional networks, opportunities to participate in the physics community beyond as a student, and the development of a stronger sense of belonging within the field due to a rekindling of interest. All of these experiences could in turn translate to fostering a stronger physics identity [41-43]. Broadly we are looking to determine *can the Community of Practice framework be utilized to understand how informal physics programs aid or dissuade students in developing a physics identity?*

---

[1] In this paper, we are using "outreach" and "informal" interchangeably to mean opportunities to engage in physics teaching outside of the formal classroom setting.

For this purpose, we operationalized the Community of Practice framework for the context of informal learning spaces to characterize the development of physics identity among the university students who facilitate them. In particular, for this paper we are focusing on answering the questions: how can we operationalize the framework to help us establish membership (and therefore identity) within the community of practice? and what structures and/or mechanisms within that community foster (or hinder) the membership development?

Our study sits between the crossroads of identity, the Community of Practice framework, and program design, but in this theoretically driven paper we are focusing on the operationalization of the CoP framework in the context university students facilitating informal physics programs. Currently, we are using CoPs as an analysis framework for understanding how identity is developed within informal physics programs. However, we believe that future work in the PER community can utilize CoPs as a design framework for the development of future programs. In this paper, we share the stories of two physics graduate students who facilitate informal physics programs. We picked these two stories, from our analyzed data set, to demonstrate the use of the CoP framework to understand how the experiences of facilitating informal physics programs can support the development of their physics identity. We identify how the CoP constructs of the informal program are manifested in the students' self-described experiences. The different operationalized layers of the CoP framework allow us to identify the university students positioning within the informal physics program CoP , the broader informal science CoP, and the physics CoP. We can also see how these communities overlap in the students' experience. Furthermore, we are able to identify what factors have a bigger influence on these identities, and if there are shifts in their identities through participation. The implications of the operationalization of the CoP framework are relevant not only for informal learning spaces but also for formal learning environments and the physics field at large. These implications can lead to an understanding of what factors have a bigger impact on supporting identity formation and can therefore lead to the design of more inclusive learning environments.

## Considering Identities in Communities of Practice

A broad definition of identity refers to the qualities and attributes that make a person (or group) who they are and is often aligned with sociocultural labels, such as gender, race, and socioeconomic status, that allow for categorization of individuals and groups [44-46]. More specifically, identity has been established as a social construct that links the individual with the social world by translating social norms to self-categories and establishing positionality of individuals and their relationship with other members of society [47-50]. Furthermore, research has determined that the individual idea of self is formed by a series of identities that interact with each other, i.e. an individual is the sum of the parts and each part has something to contribute [26, 29, 51]. For example, a person can simultaneously be a physicist, a black woman, a runner, and a mother; all of those identities overlap and depend on each other to make the individual whole. Therefore, our understanding of self and identity is dynamic, constantly subjected to change, being reassessed and molded by interactions of our individual world with social relations and collective spaces [51].

A person's discipline-based identity, such as a physics identity, represents one of these parts, and it is related to the individual's perceived association with physics [21, 52]. As a sociocultural construct, an individual's physics identity is mediated by their career interests, social environment, cultural norms, and interactions - it is how we are perceived within our disciplinary role, and what resources are available to pursue those interests [31-33, 53-55]. Therefore, the culture surrounding the discipline and the departments in which we conduct our studies can highly affect our discipline identity. Recent research indicates that a stronger connection with the discipline (i.e. a development of a discipline identity, such as a physics identity) increases the chances of pursuing and persisting in the discipline [19,21,33]. The majority of studies on science and physics identity have focused on identifying the characteristics that contribute to the formation of a science/physics identity [21, 48, 56] i.e. what are the characteristics of the individual and their experiences that will indicate the formation of a physics/science identity? However, if we define identity as a social construct and the combination of multiple identities, then we also need to understand how the parts develop and interact, what activates the different identities under different contexts, and how the collective impacts and shapes these multiple identities.

Community of Practice (CoP) [37, 51] is a social theory of learning that started from Vygostky's constructivist ideas of learning and the Legitimate Peripheral Participation framework [36, 37]. Wenger expanded the theory to incorporate four components of learning, which are meaning (the ability to experience the world), practice (shared resources, perspectives, and norms that guide our mutual engagement), community (the social configuration in which we participate and are recognized), and identity (how learning and participation define and change who we are).The framework was developed from a longitudinal study in which Wenger [51] followed a group of insurance claim processors to identify how the individuals and the collective learned. During this time, Wenger identified the common unspoken norms and practices that the individuals have adopted as part of the collective and how moving towards developing expertise in those practices grants the individual seniority within the group. It is through these norms and practices that individuals recognized themselves as part of the collective which is the community of practice of insurance processors. Through the individuals' interactions and what they did inside and outside the community, Wenger was able to build an understanding of how they defined membership in the community and what affected that membership.

Communities of practice are groups of people that together engage in a learning process and work towards achieving learning goals, but not every community represents a community of practice. In order to be identified as a community of practice, Lave and Wenger [37] defined three main characteristics: i)

The *domain* represents the set of shared interests, passions, and goals. A member of the community actively participates in activities that contribute to these common goals. The domain is connected to the vision of the community and members of the community share a set of skills and expertise necessary to achieve these goals; ii) The *community* is formed by the members who work towards the common interest and help each other achieve the CoP domain. Therefore, they engage in common activities, build relationships, exchange information and knowledge through interaction, and learn together; iii) The *practice* relates to the sets of tools, principles, norms, language, methods, and resources used to attain the CoP domain, to interact with other members, and participate in the activities. Some of the practices can be explicit, such as the use of scientific method in the physics community, others are more implicit, such as hidden curriculum in physics, related to the way we use and interpret mathematics compared to other disciplines [57]. We can consider the physics community as an example of a CoP. The physics community (even though there are subgroups based on research topics, projects, departments and societies across the world) can be viewed as a collective that is working toward a common vision to develop a deeper understanding of the universe's behavior (the CoP physics *domain*). Each of the members of the physics community make up the *community* and bring their diverse expertise to achieve that goal. Finally, there is a set of common norms and practices, such as the use of mathematics as a language to communicate nature's phenomenon and the use of scientific methods to approach the validity of a theory or hypothesis (which compiles the *practices* of the community).

The CoP framework informs how identity is created through practice as a social enterprise in the form of membership within the community. As we become more central members of the community, our identification with that community also grows and the community's values and practices become our own allowing us to therefore build an identity that is associated with that community. "Building an identity consists of negotiating the meanings of our experience of membership in social communities" [51. p.145]. There is a bidirectional connection between a Community of Practice and identity formation. By acknowledging the individual, the community is giving the individual a sense of belonging; through the practice, the individual negotiates how s/he participates in the CoP; and together the community develops and engages in the practices. This sense of belonging, negotiation of participation, and engagement in the practices impacts how a member's identity is formed within the CoP.

In physics, previous studies have used the CoP framework to study how students and teachers learn and develop physics identities [31-35]. Among the more relevant studies that informed the current study are: Close, et al. [33,34], who used the CoP framework [29] blended with a physics identity framework [7] to look at how university students' physics identity was impacted while participating in the Learning Assistance (LA) program. They used a blended framework to understand the connections between elements of identity, more specifically, they used the CoP framework to determine what were the influential factors within the program that impacted participant's identity. A different study by Irving and Sayre [35] studied physics identity development within formal learning environments by taking a close look at upper level physics courses. They used the CoP framework to determine whether or not the students are showing signs of changes in their membership in the physics community while they participate in different upper division level classes that are central to the physics curriculum [35]. These studies used the community of practice framework to study identity formation differently. The first one looked at what elements within a community of practice can promote identity development, while the second one looked at changes in the membership within the community that indicate shifts in identity development. The novel aspect of the current study is that we seek to build on Wenger's CoP framework to establish a tool that would allow us to: 1) Determine membership in the corresponding CoP; 2) Identify the mechanisms within the communities of practice that allow movements within membership levels.

# Study Design

In this theoretical paper, we operationalize the CoP framework to determine membership and mechanisms that affect identity within the context of informal physics programs. In alignment with the constructivist nature of the Community of Practice framework [36], we have chosen to use narrative inquiry [61] as a methodology to capture the meaning of the stories shared by the university students who facilitate informal physics programs. Through their narratives we can identify the important aspects of their experiences of becoming physicists and facilitating the programs and how these aspects intertwined. The overarching study is to understand how programs' design support or hinder physics identity formation. For this purpose, we conducted interviews with facilitators from two different informal physics programs. During the process of operationalization of the program we focused on 4 interviews, two from each program. We used these interviews to redefine the constructs of the CoP as well as the specific subcodes that emerged from the operationalization process. In this paper, we present how the operationalized CoP framework can be used to explore physics identity development by telling the stories of two of the physics students interviewed.

Narrative inquiry is a qualitative methodology that can be used to analyze data for restoring stories and developing the themes that appear in those by using the content of the interviews as well as the language and the context on which those stories are presented [61]. Through this methodology we can capture the complex and dynamic process of constructing and negotiating individual identity within a collective. In this work, we use the narratives to redefine the elements of the Community of Practice framework and identify the connections between the elements that define membership within the community and those that prompt movement within membership levels. Once the different structures of the framework are redefined we can use them to characterize identity formation within programs and understand the complex and dynamic nature of identity development. The idea of restructuring the framework is to have a theoretical tool that can help us establish a profile of identity/membership but also provide an insight of what practices/structures can be changed or modeled in current or future practices to promote identity/membership development.

## Collecting Interviews

In this work, we are investigating membership and identity development within the context of university students participating within educator roles in informal physics programs, which we referred to as university educators (UEs). To collect these stories, we developed a semi-structured interview protocol to learn about the UEs, their journey of becoming physicists, and what led them to engage in informal physics. Most interviews concluded with a short demographics survey. We asked questions such as: Why did you decide to volunteer for the informal program?, In what ways has participating in the informal program benefited you?, What have you gained from those experiences?, What is the most important thing you got out of them?, How did you end up in physics?, Do you identify as a physicist?, and What were the most important factors that landed you in physics and what motivates you to continue? Overall, we conducted 23 semi-structured interviews with UEs volunteering in two informal physics programs, hosted at two R1 institutions in the midwest of the United States. We chose these two programs because the design of the programs are very different but both programs utilize undergraduate and graduate students in facilitator roles. One of the programs is designed as an after-school physics program that is ongoing during a complete semester. The other program follows the format of a demo show, with a traveling van component during spring break. The UEs come from different backgrounds and degrees, ranging from undergraduate and graduate students, postdocs, and staff. Thirteen of the UEs identified as males and ten as females and all but two identified as white.

For the process of operationalizing the framework, we used purposeful sampling [62], that is we selected the cases that presented different and contrasting perspectives on their experience within physics and the informal program they facilitated. Some of the contrasting perspectives we were focusing on to select the interviews were for example, graduate students vs undergraduate students, males vs females, from different programs, levels of struggles in their career development, and physics major vs non-physics major. The main reason for this approach was to test the CoP constructs and whether it could differentiate between the experiences and program structures. From our large sample we identify four UEs that had enough variance and similarities to help us contextualize the individual constructs [39, 41, 42]. We selected two graduate physics students, female and male, both whites who participated in the after school physics program and two undergraduate students, a black female physics major and white male non-physics major, who facilitated the demo show. In this paper we are presenting the stories of the two graduate students who facilitated the afterschool program, which we will refer to as Physics Can be Awesome (PCA). We assumed that physics graduate and undergraduate students have already developed some form of physics identity and we are observing if and how it morphs through participation in informal physics programs. These interviews were selected on the basis that white males are the highest represented demographic in the physics field while women are among the underrepresented groups in physics. Furthermore, their stories are well representative of the struggles that women experience in physics, compared to white males.

We would like to make a note of the identities of the researchers and authors involved with this paper in order to address all identities for everyone involved. As with any study involving human interpretation, the identities of the researchers are important aspects of possible bias among the study design. The interviews presented here were conducted by two female physics education researchers, a Latina woman and an African American woman, who were also involved in the informal physics program. Their involvement and experience with this informal physics program allowed the two researchers to further the discussion during the interviews and ask probing questions about specific aspects of the program. Furthermore, two of the authors of the paper have been program directors of PCA, which means that they also bring a CoP central member perspective to the contextualization and operationalization process. These experiences allow the authors to provide insight that is useful in the interpretation of each interview.

## Context

In this work, we are investigating membership and identity development within the context of university students participating as the role of educator in Physics Can be Awesome (PCA). This informal physics program is funded through a physics frontier center at a large R1 institution within the United States and provides opportunities for university students, mainly graduate students, to engage with local youth in physics activities. The university students commit to a semester of weekly meetings where they co-construct physics inquiry-based activities and experiments with small groups of youth participants. Each semester, there are about 30-35 volunteer student facilitators and around 120 youth participants. The large majority of volunteers (>70%) are graduate students in their first years associated with the physics frontier center and the other volunteers are undergraduate students and postdocs. The volunteer demographics change from semester to semester but on average there is a larger percentage of males and the majority of the volunteers are white, which is consistent with physics department representation [58, 59]. All of the PCA sessions are held at middle schools local to the physics frontier center that have majority hispanic/Latinx student populations as well as >50% student participation in free/reduced-cost lunch programs.

PCA has many unique aspects that set it apart from other informal physics programs on both the facilitator and participant side. All of the physics activities and experiments are based upon the constructivist model for educational informal environments and are designed to be exploratory activities involving intergenerational work. Each group of youth participants is paired with one university student facilitator, which we refer to as a University Educator (UE), and each week the group collectively chooses what activity they would like to work on (either a continuation from the previous week or a new topic can be chosen). The weekly sessions are held during after school hours and all of the youth participants are self-selecting. As part of the program experience, each UE received research-based training in physics pedagogy as well as training that discusses science communication, curriculum content, and issues of diversity, equity, and inclusion.

In this work, we postulate that PCA is a CoP that has the potential to support the physics identities of its members. From our knowledge of the informal program (via information gathered on their website, conversations with facilitators, staff members, and the director of the program, and through observations of the activity sessions, as well as the fact that two of the authors were involved in the program at different stages), we can identify the three essential aspects of a CoP- domain, community, and practice- within PCA. When considering University Educators and youth participants as members of the CoP, we can see that all players share a domain of increasing excitement and exploration of physics through various inquiry driven activities designed for the youth, the community is formed by both the EUs facilitating the program and the youth that choose to engage weekly, and a practice that consists of a wide range of fun, hands-on physics activities. We further hypothesize that participation in the weekly sessions can be highly impactful for those that choose to volunteer as a UE. Due to the nature of the program, each group of youth and facilitators are consistent each week, which allows for a long-term development of rapport, communication, and growth to take place within the group.

We chose to present two interviewees from CPA within this paper. Both interviewees were pursuing a PhD in physics during their time in PCA and had the same role as facilitators. However, the two interviewees are at different stages in their PhD and have different levels of experience in PCA. For the sake of anonymity, we changed the names of the interviewees and removed any information that could make them identifiable. The first interview participant, Cecilia, is a female, third year PhD physics student that had just completed her first semester facilitating in PCA at the time of her interview. She has had previous experiences facilitating informal programs. Cecilia had started her PhD at an institution abroad but decided to take a different path within the first year and resumed her PhD a year later at the university where she completed her undergraduate degree. The second interview participant, Mike, is a first year male PhD physics student, who had just completed his second semester facilitating in PCA at the time of the interview. Mike had not had any experience facilitating informal physics programs prior to participating in PCA. However, unlike Cecilia, he had a straightforward path from to college and to the PhD program with no struggle in transitioning between educational phases in his life.

Throughout the remainder of the paper, we will describe how the Communities of Practice framework has been contextualized for informal physics programs. Furthermore, we will detail a case study of two University Educators who were facilitators within PCA and use the framework to understand their experiences and how CPA might have shaped their physics identity.

## Structuring the Community of Practice Framework

As it was presented by Wenger [51], the Community of Practice is a theory of learning; the framework based on constructivist ideas in which the learner forms part of a collective and it is through the negotiations and interactions with the collective that the learner acquires and assimilates knowledge and practices. The purpose of the theory is to establish a set of principles that could help understand how learning happens in social contexts and what structures support it. In this theory, Wenger [51] establishes four components that characterize learning as a social enterprise: learning by doing (practice), learning as experience (meaning), learning as belonging (community), and learning as becoming (identity). The community of practice demarks a unit of analysis for which Wenger lays out the characteristics that define it, how it evolves, and sets boundaries, while the practices are the glue of the community. Under these premises three dimensions are introduced that connect the characteristics that comprise a CoP (domain, community, and practice) with membership and are meant to help answer questions such as: What is the community about? (domain), How does it function? (community), and What are the competences developed? (practice).

When shifting the focus from the collective to the individual, the theory allows us to understand identity development through membership in a social context. Wenger characterized how the practices of a community can shape identity formation; i.e. identified a set of mechanisms and structures within the communities of practices that can foster the development of an individual's identity through participation. This formation of identity or membership is a dynamic process that is constantly being negotiated. For this reason, CoP theory explains that membership can take different trajectories which determine how the individual chooses to participate in the community [51].

The process of restructuring and operationalizing the Community of Practice constructs was iterative. In the process of restructuring the framework, we divided the constructs into three layers (shown below in Figure 1). In the first layer, we redefined and contextualized Wenger's dimensions as quantifiable constructs to help us determine in which trajectory (or level of membership) the individual perceived themselves as within the community of practice; i.e. the community dimensions become a measurement of membership in the community, which means the trajectories become subcodes of the community dimensions within our structure of coding. Subcodes were used to identify a dynamic standing or movement within membership levels for both the community dimension (layer 1 in figure 1) and the mechanisms of identity (layer 2 in figure 1): *inbound, outbound,* and *neutral*. The *inbound* subcode described when the individual is actively looking to move into a more central membership role or increase their involvement with the community. For example, this might mean returning back to the community after a break or trying to become more familiar with the values and goals of the community. The *outbound* subcode describes when the individual has decided to move out of the community or to move to a less-committed position within the community. This could be due to a negative experience within the community or due to graduation. Likewise, the *neutral* subcode is used when the coded segment does not clearly indicate in inbound or outbound movement.

We can also identify what level of membership the interviewee is describing during a specific portion of the interview. In addition to different trajectories, Wanger describes five different levels of which members of the community can participate [37, 51, 60]: *transactional* or *outsider, peripheral, occasional, active,* and *core*. The levels of participation within a CoP are dynamic and can change at any time, but there is a spectrum that goes from outsider to central membership (for static membership levels) related to our self-identification with that community (see Figure 3). In our restructuring of the framework, the subcodes that identify the level of membership demonstrated in the interviews are *insider* and *peripheral*. These two subcodes denote moments when the interviewees discuss feeling or acting as if they are on the outskirts of the community (*peripheral*) or core members of the community (*insider*). We chose to only focus on different levels of membership, each representing opposite ends of the spectrum, so that we easily define and identify these membership levels within our data. These membership level subcodes are used in conjunction with the community dimension construct (layer 1 of the coding framework seen in Figure 1).

A second layer of constructs refers to the way in which we used the characterization of identity in practice as a measurement of how engagement in the activities of the community shapes the individual's membership. These mechanisms demonstrate the dynamic nature of identity formation and how the practices the members engage in can push them in an inward or outward trajectory. To see these movements, we added a set of subcodes that would allow us to see the direction of the movement the mechanisms were pushing the membership towards. *Inbound, neutral,* and *outbound* are defined the same way here as we discussed above.

Finally, in Wenger's theory of CoP there is a discussion about the nature of boundaries and how the community and practices can sit in the intersection of broader social structures. In the process of restructuring the framework, we felt the need to delineate these boundaries by creating a third layer that would indicate which particular interest area/community(ies) were being impacted by the engagement of the individual in the practices. While the majority of the interview questions focused on the participants' involvement with their informal physics program, there were times when interviewees brought up their involvement with other communities or interests during their response.

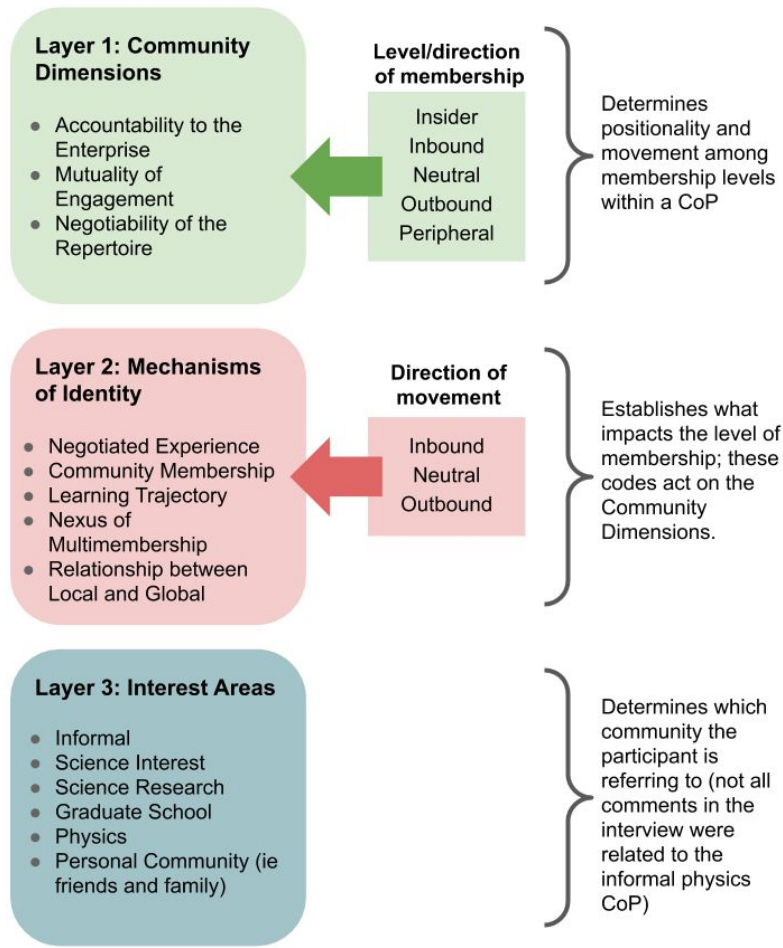

**Figure 1:** Depiction of three main layers of our operationalized CoP framework. The first layer consists of the community dimensions, the second layer consists of the mechanisms of identity, and the third layer consists of the interests (or communities) that the interviewees were referring to during the identified segment of the interview. Each segment of the interview labelled with either a construct of layer 1 or 2 will also have a subcode indicating the level of membership currently experienced by the interviewee or the direction of movement within the levels of membership.

We felt it was important to document these parts of their response because it provided insight into how their participation in other groups intersected with their experience in informal programs and physics. During the coding process, the following communities or interests emerged within the interviews: *the corresponding informal program community, graduate school, personal life* (i.e. family and friends)*, physics, science interest, science research,* and *informal interests* (i.e. involvement in other informal spaces). There was a specific category defined for grad school because the UEs who participated in that category often referred to specific practices and norms of being a graduate student that were not present in the undergraduate students' interviews. These emergent codes are a representation of the interests and communities within our data set and are not representative of all of the possible communities that could have been discussed in the interviews (for example, no participants brought up sports teams that they were a part of).

This complex framework describes how we as individuals become more central members of a community of practice and what factors influence our membership in the communities. Figure 2 shows a visualization of the different elements of the framework within our restructuration, how the community dimensions are related to the characteristics of a Community of Practice, and how the mechanisms of identity interact.

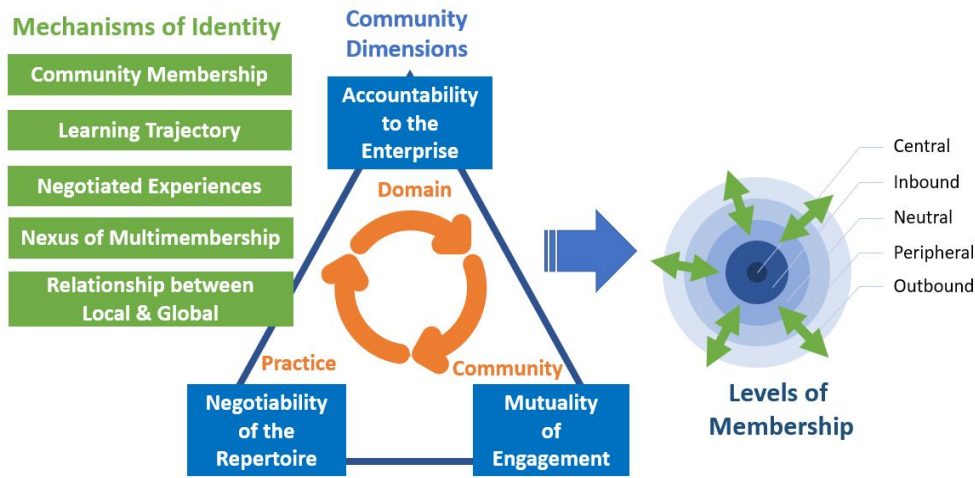

**Figure 2:** Community of Practice Framework: The identity mechanisms can shape the CoP dimensions provoking movement on the membership levels, therefore impacting an individual's identity. This image is a graphic representation of how we use the CoP (adaptation from Wenger [51]). The green arrows on the level of memberships indicate the bidirectionality of the mechanisms of identity on the community dimensions.

## Becoming a Member of a Community of Practice

We defined the first layer of the framework to determine positionality within the community of practice, i.e. a set of constructs that can help quantify the individual's perception of their membership in the community. For this purpose, we part from Wenger's [51] dimensions. Wenger determined that participation in the Community of Practice can be different depending on your level of participation and engagement with the community [60]. Participation can take many forms depending on your involvement and competence -  defined by the community as what it values through engagement and participation - within the community and its practices. The community dimensions help us determine the degree of membership of the interviewees within a particular community, based on an individual's competence and involvement in each of the dimensions.

The first community dimension is known as *Accountability to the Enterprise* and it is related to what the community is about. Refers to the understanding, valuing, and working towards the community *domain* by the members of the CoP. Lave and Wenger define Accountability to the Enterprise as a way to consider how members understand the goals/objectives of the community, how they take responsibility and embrace those goals, how they contribute to the achievement of those goals, and their constant negotiation and redefinition of the goals. Because the mission of a community is a joint pursuit, it creates a liability and responsibility among the participants, which in turn defines the practices of the community. This does not mean that all members of the community hold the same ideas or opinions, but that they collectively work towards that higher goal. For example, in the field of neutrino oscillation physics some experiments have confirmed the existence of sterile neutrinos, while others have not. However, these experiments have a common goal which is the study of neutrino oscillations. It is in this constant negotiation and accountability that the community evolves. The Accountability to the Enterprise code describes the level of commitment and group involvement of individual participants and shows how a member's understanding of the objectives evolve over time.

Table 1 shows quotes coded as Accountability to the Enterprise due to the interviewee explaining their views on the importance of PCA for the program participants (the insider quote), their personal reasoning for joining PCA (the inbound quote), the idea of PCA being a time commitment (both the neutral and outbound quotes), and a description of previous experience in PCA (the peripheral quote which shows the interviewee as a peripheral member of PCA in the past). Each of the inbound and outbound quotes show movement (either into or out of the community) because the interviewee is discussing the reasons for joining PCA (showing inbound movement) or reasons for why they may not continue with involvement (outbound movement). In addition, both the insider and peripheral quotes show the interviewee as being very involved with the community and thoroughly understanding the goals (insider member) or as a peripheral member of the community who does lab tours but does not go to the schools for the afterschool programming.

**Table 1:** Examples and explanations for each subcode under the Accountability to the Enterprise construct.

| Membership Code | Subcode | Examples |
|---|---|---|
| **Accountability to the Enterprise (AE)** Describes how members of informal physics programs perceive program goals, the impact of the program, and what is important for them to be a facilitator. As well as how committed the facilitators are to the domain and therefore participating in the program. | Insider | "*The thing that I hope to convey to them is, you know, how they can do it, how good they are basically at doing science*" <br> This was coded as insider because Mike is describing what his objective for being in PCA is, which it is also the PCA mission but he has internalized it as his own in this quote. |
| | Inbound | "*[PCA] just seemed like the most applicable outreach opportunity. I loved the fact that it was going to schools and doing physics. Yeah, I think it was really the main one that I saw in science for outreach.*" <br> In this case Mike is expressing his initial interest in joining PCA, becoming a member of the community. PCA's domain aligns with his interest in physics. |
| | Neutral | "*Yeah, this is the first semester I've done [PCA]*" <br> We coded as neutral because it was not possible for us to discern no previous participation or future had any effect on membership and accountability |
| | Outbound | "*Yeah, and if I had known how crazy the first semester would be, I probably wouldn't have done [PCA]...*" <br> Coded as outbound because the time commitment between PCA and other communities was a factor in possibly pushing member out of the community |
| | Peripheral | "*Because I've given actually- Okay, to back up, I've given lab tours for [PCA] in the past, but I never actually did [PCA].*" <br> Indicated as peripheral because Cecilia is mentioning that previously she had participated in some PCA activities (given lab tours) but was not a member of PCA at that time. |

The second community dimension, *Mutuality of Engagement,* is related to how the community functions and the forms of interactions between its members. Members of the community participate in activities and efforts that contribute to the community's domain, and the way in which participants engage in those activities is constantly negotiated and redefined. It is the understanding of these actions that defines members and differentiates them from non-members. Within a CoP everyone has their individuality that contributes to the community's engagement and practices; some of the roles and contributions may overlap, and others might be complementary, but all add to the development of the community. For example, a particle physicist studies the interactions of the fundamental blocks of matter, while a nuclear physicist studies nuclei interactions. There is a large overlap between these two members of the physics community of practice since they are studying matter interactions, just at different energy and substructure levels. While they share similar practices and languages, they have different expertise, which both contribute to the advancement of knowledge of the universe (domain of the physics community at large).

**Table 2:** Examples and explanations for each subcode under the Mutuality of Engagement construct.

| **Membership Code** | **Subcode** | **Examples** |
|---|---|---|
| **Mutuality of Engagement (ME)**<br>Describes the interactions the interviewee has with other members of the informal physics program. For example, the interactions could be with peers, audience members, or directors/coordinators. This code can also describe how the individual is recognized by other members within the community or how other members of the community influence that individual's involvement. | Insider | No specific examples for Cecilia & Mike |
| | Inbound | "*[O]ne of those kids just straight up told me I was his favorite in one of these, and it was adorable because he's just, the kid's really sharp... [H]e was my favorite. It was nice when you kept coming back because they, you know, they know your name and they'd be like 'hey it's you, work with us!' you know?* "<br>Cecilia is describing how she was building the rapport with the children and as she kept coming back those children will recognize her as one of their group, therefore helping her move inward to the community, developing a sense of belonging |
| | Neutral | "*[T]hey weren't as involved, I guess, it seemed like. You know, high schoolers are not as receptive to making new friendships or connections or that sort of thing with grad students or whatever.*"<br>Coded neutral because we could not discern if the fact that the highschoolers were not engaging as much with him or the activities was detrimental or beneficial to his membership in PCA. |
| | Outbound | "*Oh yeah, yeah. Sometimes it was really difficult to kind of get them oriented towards writing in their journals or notebooks, and sometimes they just wouldn't listen...*"<br>Coded as outbound because Mike is describing something that was not a positive experience, which could possibly impact his membership. |
| | Peripheral | No specific examples for Cecilia & Mike. |

Table 2 shows examples of Mutuality of Engagement with our interviews from Mike and Cecilia. There were examples from each of the three trajectories within Mutuality of Engagement for Mike and Cecilia, but there were no specific examples of the insider or peripheral membership levels within these two interviews. However, an insider Mutuality of Engagement code would look like a description of norms of interaction between participants of the community, what they talk about, how they talk about it, other types of communication or interactions. Likewise, a peripheral Mutuality of Engagement code would look like a description of an interaction with other members of a community in which norms and practices were not clear, just testing out participating in the community, or a case where members of that particular community do not perceive you as a full member of the community. Both the inbound and outbound trajectories are relating to how the facilitator interacted with the children participating in the program. Positive interactions with the participants seemed to be very important to the facilitators (inbound) while negative or difficult situations with the participants lead to some frustration (outbound).

**Table 3:** Examples and explanations for each subcode under the Negotiation of the Repertoire construct.

| Membership Code | Subcode | Examples |
|---|---|---|
| **Negotiation of the Repertoire (NR)** Describes the practices specific to the informal physics program. This includes descriptions of the specific activities or demos that take place during the informal physics program and knowledge that facilitators need in order to reach the program's goal. This code would also highlight when the individual describes their perception of the skills they may or may not have. | Insider | "*I mean the kids, you know, especially again as you, like, build a rapport with them, they were kind of- you could kind of just be like 'listen, this is cool, this is what we do. Like you know, I could give them advice. They'd be like 'hey I want to do something neat,' and I'd be like 'the spectroscopy experiments! My favorite because you get to look at cool colors!' or 'you should play laser chess because laser chess is awesome!' and they'd just be like 'okay!*" Here Cecilia is talking about the PCA practices as hers as well as being confident about sharing those with the children. |
| | Inbound | "*What I've gotten out of it is just being a better teacher, that's for sure, having a better understanding of- but also a better sense of what schools are like these days and what kids are like these days; you know, what it takes to get them into science, kind of. I'm generating a sense of that.*" Here Mike expresses that he has and is still learning the PCA practices, therefore a dynamic membership moving in. |
| | Neutral | "*It wouldn't be a deterrent for me to do PCA, or to do it at that school. But yeah, I would just want to learn a little more about what the concept is behind it.*" Coded neutral because Mike is talking about some new practice of PCA that he is unsure about whether he would engage in it or not. |
| | Outbound | [Physics Community[2]] "*No, especially when I started grad school or like when things got really, really hard or I was struggling with, like, an experiment or something, like clearly, I just don't belong here.*" Coded as outbound because there were instances in which Cecilia found it difficult to engage with the practices of the physics community and made her question her membership in it. |
| | Peripheral | No specific examples for Cecilia & Mike |

---

[2] There were no Negotiability to the Repertoire-outbound codes for PCA community, so we used one from the physics community to give an example

The final community dimension, *Negotiability of the Repertoire,* relates to the competence developed while participating in the CoP. This dimension refers to the understanding, learning, and participating in the practices of the community. It encompasses the set of resources, routines, language, methods of carrying out actions, symbols, and concepts the community has developed through the negotiation of meaning and on working towards the community's goals. In the physics community, this can look like sets of laws, mathematical language, or methodologies that we use to pursue the understanding of the universe.

Table 3 shows examples of Negotiation of the Repertoire from Mike and Cecilia's interviews. We have provided examples related to the PCA for the insider, inbound, and neutral subcodes. The outbound subcode did not have any examples for the PCA community, but we did provide an example from the physics community. There were also no specific examples to provide for the peripheral membership level. However, a peripheral code would look like a description of an experience that indicates the individual is not fully familiar with the practices or ways of doing things, an indication of not feeling comfortable or interested in engaging in the practices, or instant of being told by other members of the program that they are engaging in the practices appropriately.

## Forming Identity in a Community of Practice: Mechanisms of Identity

The amount of competence an individual has in a community is related to their community membership, and therefore competence influences identity development. In this study, we are interested in what within the community of practice can foster identity development. Therefore, the second level of our framework is focused on identifying what mechanisms within the informal physics community influence identity changes among the University Educators who are facilitating the program. Furthermore, identity is a dynamic construct and therefore the level of participation in a community is likely to change or fluctuate because different environments and experiences affect our engagement with the community. This is why we use the *inbound*, *neutral,* and *outbound* subcodes with each mechanism of identity.

To determine what influences these changes, the CoP framework identifies a set of factors or mechanisms (five total) that can impact a member's level of participation in the community; therefore the mechanisms foster or suppress an individual's sense of identity within that community [51, 60]. The first mechanism of identity development is known as *Negotiated Experience*. This construct captures the process of making meaning of experiences through participation in the community, including interactions with other members, and how those experiences form the individual's perceptions of themselves as members of the CoP. Table 4 provides examples from our two interviews for each trajectory subcode for Negotiated Experience. All of these examples are related to how the UEs interacted with the student participants in PCA; some interactions were positive and made the UE feel like the students were building a community and wanting the UE to be involved (inbound example). Other interactions were not as positive causing the UE to become frustrated (outbound example).

**Table 4:** Examples and explanations for each subcode under the Negotiated Experiences construct.

| Mechanism Code | Subcodes | Examples |
|---|---|---|
| **Negotiated Experiences (NE)** Captures how UEs make meaning of experiences that they have through participation in the informal physics program and by interacting with other members of the CoP. Through these experiences and interactions with others, the UEs define what is valued. | Inbound | "*But when I worked with Pedro and Luis, I felt like [...] I would come and there would be like a reunion, like 'hey guys', and they'd be like 'hey!' [...] it was cute. Like when they came by the lab [..] I think Pedro asked if he could work at my lab when he got older or something. It was just pretty fun. Yeah, I definitely felt like an older brotherly connection with them or something.*" This relationship between Mike and those children came out several times throughout the interview and was definitely a relationship that helped Mike move his membership inward, feeling more connected to the PCA community. |
| | Neutral | "*Some days I definitely came away a little frustrated, but you know, that never translated to the interaction with the kids or anything. It was always fun, just maintained a fun atmosphere.*" Coded neutral because it was not possible for us to discern if the frustration or the fun induced any particular movement within the membership. The frustration was due to children not engaging with him on the activities. |
| | Outbound | "*[I] didn't expect how frustrating it could be sometimes when the kids just aren't listening to you.*" Some interactions Mike experienced with the children were sometimes detrimental to his participation in the community. |

The second mechanism of identity, *Learning Trajectory*, is a construct related to things that have been learned which resulted in the participant becoming a member of the community. This code incorporates past and possible futures into making meaning of the present; i.e. it captures the experiences that have led facilitators to participate in different ways within the CoP. The learning trajectory of an individual influenced which elements of participation are perceived as important and which are marginal. This mechanism helps capture the experiences that give individuals context to determine what things are (or not important) and what has been learned along the trajectory. Table 5 includes examples from our two interviews for Learning Trajectory. However, it is important to note that the neutral and outbound examples are not related to the PCA community, but rather are examples from when the UE was discussing their desire to do physics and their experience within graduate school (outbound example).

**Table 5:** Examples and explanations for each subcode under the Learning Trajectory construct.

| Mechanism Code | Subcodes | Examples |
|---|---|---|
| **Learning Trajectory (LT)** Captures what past experiences lead the UEs to participate in the informal physics program and why they might value certain experiences. In addition, this code would also capture possible future actions that the UEs discuss based upon their past and present participation in the community. | Inbound | *"[I] didn't realize that being a scientist was a thing until I was like sixteen, and I would've loved it somebody came, like you know, when I was eight years old and was like 'hey, look, lasers are cool, you can do this!'"* When asked about why she joined PCA Cecilia talks about her past experience as a motivator for participation. |
| | Neutral | [discipline community[3]] *"So yeah, I wanted to do biology at first. I kind of realized that- the more I thought about it, I realized that the questions you ask and answer in biology aren't the kind I'm necessarily that interested in."* When asked about how he ended up in physics, Mike discusses his initial interest in biology. But it is not clear - from this quote - if the realization that it was not biology was what pushed him into the physics path instead. |
| | Outbound | [graduate school community[4]] *"One of the interesting things was I was actually a grad student for like a month and a half in Innsbruck, Austria. [...] that ended up not being like a good place for me to be at that point in time in my life, blah blah blah, you know. Short story long, I ended up dropping out."* In this case Cecilia is talking about some past experience in graduate school that pushed her out of that community (temporarily). |

The next mechanism of identity, *Nexus of multimembership,* captures how individuals negotiate being members of different communities. Each individual is composed of multiple identities and how one negotiates membership to these different communities can impact participation in the communities. This code is meant to capture all forms of participation that contribute to the complete mesh of identities within an individual. In the case of our interviews, this code captures instances where the UE is describing how participation in PCA might overlap with other areas of their life such as their physics studies or graduate school experiences. Table 6 shows examples of Nexus of Multimembership from Mike and Cecilia that displayed an inbound or neutral trajectory among community memberships but there were no specific examples of an outbound trajectory. However, an outbound trajectory for Nexus of Multimembership would consist of the interviewee describing how being a member of two (or more) communities is difficult or clashes (the multiple memberships is an inconvenience or is incompatible in some way).

**Table 6:** Examples and explanations for each subcode under the Nexus of Multiemembership construct.

---

[3] There were no Learning Trajectory-neutral codes for PCA, so we used an example from the physics community instead.

[4] There were no Learning Trajectory-outbound codes for PCA, so we used an example from the graduate school community instead.

| **Mechanism Code** | **Subcodes** | **Examples** |
|---|---|---|
| **Nexus of Multimembership (NM)** Captures how the UEs describe being members of two or more communities (i.e. PCA and physics) in order to understand how different memberships contribute to their overall identity. | **Inbound** | *"[T]he younger brother is the kind of person we're trying to get through to it seems like. Clearly, he has the aptitude, and maybe if we can show him that then he would be more interested in doing it."* <br> Mike as a member of the physics community is expressing how he is trying to attract PCA children to physics. |
| | **Neutral** | *"My only real hesitations weren't related to PCA necessarily, they were more related to, you know, being able to escape the lab. And that's really just a laboratory politics thing, that really doesn't have much to do with PCA."* <br> In this case it was unclear if her participation in the graduate school/physics community was impacting Cecilia's participation within PCA. |
| | **Outbound** | No specific examples for Cecilia & Mike. |

*Community Membership* is the fourth mechanism of identity construct and is related to the proficiencies developed and valued by participants in a community. These proficiencies could be skills, capabilities, ways in which community members interact, perspectives and interpretations members share, or the use of a shared repertoire and resources. Community Membership captures how members look at the world, how they relate to others, and their knowledge of how to participate within the community. The more central a member of the community becomes, the more they are perceived as competent by other members and made to feel competent and able to perform well in the practices of the community. Thus, this code is also related to recognition received from other members of the community. Table 7 below gives examples for inbound and neutral trajectories from Mike and Cecilia's interviews. The inbound example shows an instance where the UE was participating in leading a lab tour for PCA and had a positive interaction with one of the present teachers. While there were no specific examples for Cecilia and Mike for an outbound trajectory, it would look like a description of an experience where the UE did not feel comfortable engaging in the practices or with the norms of interaction of the community.

**Table 7:** Examples and explanations for each subcode under the Community Membership construct.

| Mechanism Code | Subcodes | Examples |
|---|---|---|
| **Community Membership (CM)** Captures how the UEs are interacting with the practices of the community and how other members of the community might help further competence within those practices. This is related to feeling like a competent member of the community and understanding what is needed to participate. | Inbound | *"There were a few different tours, and one of them had their teacher with them, their science teacher, and he was really helpful in- it was really informative to me to see how he took what I said and explained it to them. I was trying to make it accessible, but he really knew how to do that, so that was cool."* Interaction with other members in the community are helping Mike learn the PCA practices and move inward in the community. |
| | Neutral | *"Yeah, and kind of like what the message we're trying to convey is, which is maybe confidence and science and arts, arts and science."* Mike is trying to explain what PCA activities are communicating and he is not completely sure he knows what is the message. |
| | Outbound | [Physics community[5]] *"[I] think a lot of my issues in kind of developing into a physicist were because I didn't quite understand what I needed to develop."* Here Cecilia is expressing that her lack of understanding of the physics practices was one of the causes that pushed her out of the community initially. |

The final mechanism of identity, *relationship between the local and global*, is related to the constant negotiation of the local ways of belonging and how that fits in a broader spectrum of practices and norms. That is, how being a member of a local CoP is connected to being a member of the more universal community. For example, this code would capture the complexities of being part of the physics community at a local institution/department and belonging to the community of physics at large. In our local CoP we engage in pursuing the domain but also figuring out how our engagement and participation fits in the broader scheme of things, fits the purpose of achieving the goals of the broader CoP. Table 8 contains examples of Relationship between Local and Global for Inbound and Neutral trajectories from Mike and Cecilia's interviews. There is a recognition that even though they are at a university, there are still opportunities to interact with and impact elementary and middle school students (inbound examples) and one of the UEs mentions their childhood environment and experiences as a way of contextualizing the student participants in PCA (the neutral example). There were no examples of an outbound trajectory for this construct within the interviews, but a possible outbound code would describe how the interviewee manages being a member of the local community but does not fit into the larger community. For example, enjoying being a member of PCA but not outreach at large.

**Table 8:** Examples and explanations for each subcode under the Relationship between Local and Global construct.

---

[5] There were no Learning Trajectory-neutral codes for PCA, so we used an example from the physics community instead.

| Mechanism Code | Subcodes | Examples |
|---|---|---|
| **Relationship between Local and Global (RbLG)** Captures how UEs are negotiating what their membership in an informal physics program means in the broader context of their lived experiences within this world. | Inbound | "[T]he school that we were at was super focused on STEM. Which is good, you know. I think that, you know, that's not to say that like arts are bad or anything, but just to have that and to have that as an option, I think that was kind of what I wanted. Like if I could just take every kid in America and just be like 'you can be a scientist if you think this is cool and you're willing to work hard,' and like, you know. So I could see a couple of them going into it.."  Cecilia is reflecting about the local situation with the school and the program and more broadly about what she would like to achieve with public engagement. Inward because her language reflects continuous participation. |
| | Neutral | "I could see a couple of them, definitely. You know, if I were to hazard a guess I'd say that most of them probably won't. But that's kind of the same thing- you know, the people that I took AP physics with like most of them are not in science, like it's just kind of how things go."  Reflecting on whether participation in PCA will lead to all the children engaging in physics in the future, Cecilia connects to her previous experiences with her peers. But it is unclear if the outcome (not every child would be a physicist or interested in physics) is something that will deter her from continuing participation in PCA. |
| | Outbound | No specific examples for Cecilia & Mike. |

## Coding Process

During this work, we engaged in an iterative coding process focusing on the two layers of the framework - the community dimensions and the mechanisms of identity constructs - in order to establish how and what affects physics identity formation within informal physics programs. In order to analyze the interviews, the audio recordings were transcribed through a paid service. We started with Wegner [51] definitions of the constructs, looking separately at the membership codes and mechanism codes, refining definitions and validating the operationalized codes. Throughout this process we also use emergent codes to create a set of subcodes that would identify the community and/or interest that the CoP construct was related to and/or impacting. Similarly, we felt the need to create codes that could designate the positionality and dynamism of the community dimensions and the mechanisms of identity. Two researchers independently coded the four interviews using MaxQDA software, first with the community dimensions and then the mechanisms of identity. The coding process was discussed and validated with a third researcher. Once consensus was reached between the three researchers the definitions of the constructs were further refined to communicate the intent of the code. To guarantee that the final operationalized codes were clear and able to be used by researchers, interrater reliability was conducted at each stage of the process. First, two researchers coded the interviews independently, identifying and resolving discrepancies along the way. Then the third researcher independently coded and compared with the other two researchers to guarantee that kappa values were larger than 0.8 was on all the codes. This process was repeated on all four interviews and was further validated by four other researchers while applied to the larger sample. Additional refinements were made based on their feedback.

See Figure 3 for an example of coded text with the membership codes (and membership level/movement subcodes), identity codes (and membership level subcodes), and corresponding communities (in this case it is both PCA and physics for both quotes). This example of coded segments from the interviews demonstrates how we layered our coding scheme when analyzing the data.

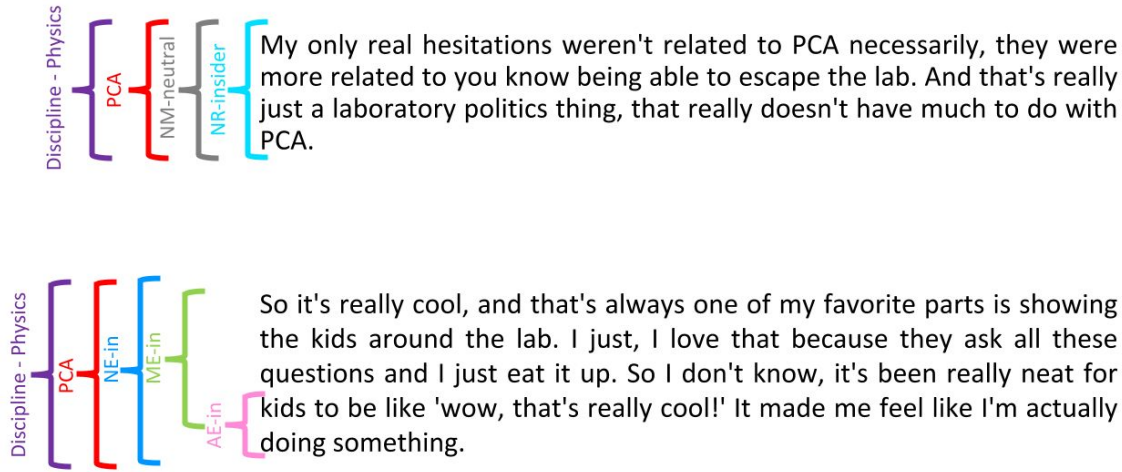

**Figure 3**: Example of coded segments from interviews. The top example shows that the text was coded as Negotiation of the Repertoire (NR; community dimension code) and Nexus of Multimembership (NM; mechanism of identity code) for both the Physics and PCA communities. The bottom example shows the first sentence coded as Mutuality of Engagement (ME; community dimension code) and the last sentences as Accountability to the Enterprise (AE; community dimension code). The entire segment was coded as Negotiated Experiences (NE; mechanism of identity code) for both the physics and PCA communities.

## The Coded Narrative

In the following sections we present the stories told by Cecilia and Mike. We focused on the content and language of the stories through the lens of the operationalized CoP framework in order to identify the important experiences that facilitated their identity formation. We use their narratives in the semi-structured interviews to demonstrate how the community dimensions and identity mechanisms are present in their stories and how those constructs interact. We identified sections of their stories in which the different elements of the framework and the connection between their participation in PCA and physics communities is apparent. Through these narratives, we see how the combination of the community dimensions and identity mechanisms tell a story about the UEs identity formation. We aim to demonstrate how the operationalized CoP framework can pick up the differences and similarities in Cecilia and Mike's stories. First, we present Cecilia and Mike's positionalities as it pertains to physics in order to help the reader understand the UEs' relationship with their physics identity. Then, we present their narrative in the context of the CoP framework to reflect on experiences, particularly as facilitators of PCA, that capture the impact on their membership within the PCA and physics community and the mechanisms that may have facilitated changes.

## Cecilia

Cecilia is a third-year graduate student in physics and has always felt as though she was interested in math and used to tinker with things at home, but she was unsure if science was for her because of the experience she had had at school. In eighth grade she was told not to believe in evolution, which she did not like, and she did not enjoy her ninth-grade biology class. However, her attitude changed when she was able to take chemistry, and later, physics. She recalls, "[I] took physics and I was like 'just kidding, I'm going to be a physicist.' And then yeah, it's kind of- I never stopped. I was one of those really lucky kids that went into college knowing exactly what they wanted to study, and I studied it, and then I went into grad school knowing exactly what I wanted to study, and I studied it."

When asked if she identifies as a physicist, Cecilia says yes, but notes that this has not always been the case. In fact, this was not the case when she started graduate school and dropped out before the first year ended. She mentions that she had some role models and encouragement along the way, including both her AP Chemistry and AP Physics teachers, that were influential in her belief that she could pursue physics. "I would say that those two were, you know, the reason why I'm here. They'll both get invited to my defense." During the interview, Cecilia also reflected on how others might identify her by saying that those outside of the field most likely see her as a physicist, but her advisors are more likely to see her as a "young physicist [...] budding sort of in the field." Overall, Cecilia feels confident in her position as a physicist and others view her as a competent member of the physics community.

## Mike

Mike is a first-year physics graduate student. He says that his interest in science started early on as a curious kid tinkering around with things, but that it was his parents' support, and in particular his mom's encouragement, that made all the difference. At an early age, Mike's mom taught him how to read, which gave him a lot of confidence to push himself and in his words "put me on a high trajectory." For example, as a 4th grader doing math in school, Mike wanted to push himself harder. With the support from his teacher, he got to work on some basic algebra problems which reinforced his confidence and ability to do things that seemed harder.

During high school, Mike initially thought he was going to be a biologist because he found his biology teacher very inspiring and he liked the lifestyle that came with being a field biologist. However, his cousin and uncle gave him some interesting books that discussed questions about fundamental science, the universe, and how it works. It was reading these books that made him realize that physics was more the sort of thing he was interested in and he went to college with the idea of pursuing a degree in physics.

Mike thinks that regardless of what triggered his initial interest or what was the catalyst that pushed him to pursue physics as a career, he would have arrived at the same place, physics. He does not doubt his position in physics and his membership within that community. He considers himself a member of the community of science/physics, even explicitly saying "we the science community" at one point during his interview. There seems to be no doubt in his mind that he was meant to pursue physics as a career and that he is perfectly capable of being a physicist. He truly feels as though he belongs in physics, but he also wants others to have similar experiences and confidence, which is what has led him to outreach.

From their statements, both Cecilia and Mike identify themselves as *insider* members of the physics Community of Practice. However, their perception of what constitutes an insider member of that community and their path to get there are different. The factors that have contributed to their perception as core members are also different. In the following sections we will present how the elements of the CoP framework help us distill those differences, as well as indicate in which form facilitating informal physics programs has contributed to their physics membership.

## Community Dimensions of Membership

### A. Cecilia

#### Accountability to the Enterprise

In her third year as a graduate student, Cecilia felt she could take the time to participate in outreach activities and says, "I thankfully got to the point where I could escape, and my advisor wouldn't kill me." She believes that outreach is important and also "looks really good" for those who are aiming to be a professional scientist. However, Cecilia also explains that she has a personal belief about educating others about science. When asked why she participates in PCA, she says "I've always thought it was important to educate people, especially kids. My reason for that is really that if, you know, I didn't realize that being a scientist was a thing until I was like sixteen, and I would've loved it somebody came, like you know, when I was eight years old and was like 'Hey, look, lasers are cool, you can do this!'" This quote was coded as Accountability to the Enterprise for the informal community with positive (inbound) movement because Cecilia is recognizing why she wants to be involved with science outreach and that it would have made a personal impact in her own life if someone else would have brought science outreach to her. Furthermore, Cecilia had participated in some outreach activities prior to PCA when she was a student abroad and mentioned, "[W]hen I came back I wanted to get back into [outreach...] So when I finally had the chance and the guts to finally say 'okay I'm going to do this once a week,' you know, that was cool." This comment from Cecilia was coded as Accountability to the Enterprise with inbound movement for the informal community because she was actively seeking to get involved with the community and explains how she took the step to become a member. This understanding of personal reasons for doing outreach and the commitment needed for outreach demonstrates Cecilia's commitment to PCA, but also to the outreach community more broadly. When she was not participating in outreach, Cecilia says she missed it and wanted to participate again because she "kind of [has] this obligation to help educate" (insider for the informal community) and therefore she believes that outreach will be part of her career in some form or another.

In this part of the story, Cecilia describes her motivation and values related to educating others, which align with PCA domain. This alignment establishes a path for membership, not only in the PCA community but also in the physics community. Through PCA, not only is Cecilia fulfilling her need to help educate others but also engaging in physics concepts with other members. It is possible that this novel way of engaging with physics allows Cecilia to reinforce her membership as a physicist. Participation in PCA, believing in the domain of the community, and wanting to be a member of the outreach community as a whole in order to help others participate in physics has helped Cecilia reshape and reconsider what it means to be a physicist. This theme of reshaping her definition of a physicist is explored more in the other constructs within the CoP framework.

### Mutuality of Engagement

Cecilia started to weekly participate in PCA during her third year of graduate school. In these weekly sessions, she often worked with the same group of 3-4 children engaging in inquiry-based physics activities, such as learning about reflection while playing laser chess. While she did not know what to expect when she first started facilitating in PCA, she did find the enthusiasm displayed by the children to be a welcome surprise. Throughout her time in PCA, Cecilia made some connections with the children in her group. She mentions that it was fulfilling to encourage the children to follow their passion:

> Yeah, there was this one kid- just straight up told me I was his favorite in one of these, and it was adorable because he's just, the kid's really sharp... So, you know, I would talk to him a little bit. You know, it got a little bit off task, but I talked to him a little bit about math and I told him once that he'd love calculus. You know, I just was like- I basically told him like whatever happens, just follow what you- because you seem to really enjoy this, follow it and good things will happen for you.

This interaction was coded as Mutuality of Engagement with an inbound trajectory for PCA because Cecilia is creating joyful memories with other members in the community and is engaging with these members in ways that are not fully outlined by PCA (ie talking about calculus). These interactions were impactful to her because she developed bonds with other members of the community, allowing her to move towards a more central membership role. Furthermore, interactions like these helped her develop a sense of belonging in the PCA community: "It was nice when you kept coming back because they, you know, they know your name and they'd be like 'Hey it's you, work with us!' you know?" (inbound for PCA). This recognition by other members in the community helped Cecilia feel valued and like a part of the community and encouraged her to continue participating in the community. These interactions also helped her strengthen her membership in the physics community by reinvigorating her passion for the physics community domain. For example, when mentioning that she loved giving lab tours to the children because of all of the questions that they tend to ask during the tours, she shares that this allows her to share more about her research. She also has this sentiment reinforced by other members of the community and shares, "You know, just kind of one of the girls told me I had the coolest job ever, which again, like when you're drudging through grad school is really fun to hear. [laughter] You're like 'oh yeah, I totally do!'" (inbound for graduate school community).

These components of Cecilia's story highlight how interactions between members of the PCA community not only impacted her membership within the PCA community and the informal/outreach community at large but also her physics identity. Through sharing her knowledge and passion for physics with the children she is able to reaffirm her commitment to the physics community.

### Negotiation of the Repertoire

In her story, Cecilia has expressed her understanding and capabilities related to the practices of the physics community and how competence and confidence within those practices are key to becoming a more central member. She even describes how throughout her path to become a physicist, she did not think she belonged in the physics community because she was not good in the practices . For example, when asked if she felt like she belonged in physics she replied "*No, especially when I started grad school or like when things got really, really hard or I was struggling with like an experiment or something, like clearly I just don't belong here.*" (NR-outbound discipline). This instance of examining her identity as a physicist was coded as Negotiability to the Repertoire-outbound for the reason that she is struggling with her sense of belonging because of feeling inadequate while engaging in the practices of the physics community. However, she goes on to say "*and then of course I got the experiment working and it was kind of like what? That felt nice.*" (NR-inbound discipline), which was coded Negotiability to the Repertoire-inbound because when she regained confidence in her ability to engage in the practices, her sense of belonging - at that point - was not questioned.

Through her participation in PCA, Cecilia has learned to appreciate the spectrum of people that can be considered members of the physics community and that people do not have to be a physicist to be interested in physics. She believes that programs like PCA can help everyone understand that physics is awesome and can be fun, even if they do not pursue it as a career. In a way, participating in PCA has broadened her perception of the repertoire and practices that determine your membership within the physics community. When she is asked about what was the most important thing she has gain from participation in PCA, Cecilia says, "And so, you know, I'd say kind of the best thing that I've taken away is an appreciation for kind of the spectrum of people, and that not everybody has to be you know a scientist to appreciate science. And you know, I think that's something that I really like" (Negotiation of the Repertoire on an inbound trajectory in PCA and the physics community).

Finally, engaging in the practices of PCA has given Cecilia the opportunity to share her membership in the physics community by normalizing the practices and norms of the community and inviting children to participate in it. She recalls her interactions with the youth participants in PCA by saying, "But you know, as a whole you could kind of build a rapport and in a sense be like 'I'm a normal human being, kind of nerdy, but normal. And you can too.'" Cecilia's ideas about building up a rapport with the students has been coded as Negotiation of the Repertoire for an Insider members of PCA because she is commenting on her experience with the sort of actions that are possible during participation in PCA and her ability to build a positive rapport.

For Cecilia, her membership in PCA is deeply connected to her membership in the physics community because she believes that outreach is a really important part of being a physicist. This connection is mainly due to her belief in the importance of educating others, especially children, about science so that they can see science as a valid career option. Furthermore, she also believes strongly in instilling in children the confidence necessary to enjoy science, even if they do not pursue science as a career. Engaging in the practices of PCA gave Cecilia the opportunity to share her interests and passion from the physics community to others who may not yet be part of that community. This sharing of passion and interest potentially help strengthen her membership in the physics community.

## B. Mike

### Accountability to the Enterprise

As a graduate student, Mike started to get involved with some outreach activities because he wanted to help to get people into science. Mike was actively looking to engage in some informal activities when he found PCA (see Accountability to the Enterprise, Inbound example in Table 1). The fact that he was actively looking for outreach and that he identified PCA as the best option indicates a level of commitment to the informal community and to the PCA community in particular. However, Mike's statement about searching for an opportunity that would allow him to go to schools and discuss physics also indicates a level of accountability to the physics community because he was excited about the physics aspect of PCA. Part of Mike's goals for getting involved in informal physics was to get people into science and he identified the PCA domain as being aligned with his vision.

Furthermore, Mike expressed that his desire to get involved with the PCA community was connected to his professional growth. As he grew as a physicist, Mike felt the need to help others pursue science and become engaged with science, indicating a strong commitment to the science and physics community in particular. Mike shares this vision when he said, "[K]ind of the more I've grown as a scientist, the more I've wanted to help others get into science." This statement was coded as Accountability to the Enterprise for an insider membership level within the informal community because Mike is recognizing the informal domain as wanting to "get others into science" and is talking about how he has consistently grown to be aligned with that domain. The most important role as a member of PCA, for Mike, is to convey to children the confidence or push they need to believe that they can do science. He sees his own experiences reflected in the goal of PCA because he knows that an external push is what got him to pursue physics. The mission of PCA, as he sees it, aligns with his own values and this alignment is possibly a reason for his commitment to the PCA community.

### Mutuality of Engagement

The relationships and connections with other members are very important for Mike to engage and become a member of that community. The focus on relationships and connections was especially true for the PCA community. When looking for outreach opportunities, something that contributed to Mike's decision was the fact that he got the sense that PCA was a good organization. He recalls that "everyone was really well organized, and I felt like the kids got a lot out of it. A good group of people." This was coded as Mutuality of Engagement on an inbound trajectory for PCA because Mike is explaining how his interactions with other members have been positive and fulfilling. Furthermore, Mike often mentioned that the connections made in PCA were one of the main things he gained from participating in the program. Mike describes the interactions with his PCA peers as "cathartic" and as "a good break from the lab work." The time spent and interaction with just the other UEs in PCA were seen by him as stress relievers: "[I] mean the car trip's always just fun, just shooting the s[], and then when you're there you're doing fun activities. I usually forgot about whatever thing I was stressed about by the time I was done, which was nice." This recollection of interactions with other facilitators was coded as Mutuality of Engagement, inbound, for PCA because Mike is really honing in on how he enjoys spending time with others involved in PCA.

Along with talking about his fellow facilitators, Mike also extensively discusses the deep connections he made with children in the program, which in some cases were so important that he referred to them as a "brotherly connection." These connections are also reflected in his comments about seeing the children engage with the activities, particularly when they came to his lab and he had the opportunity to share his research, as shown in the example given in Table 4. This quote from Mike's interview was coded as Mutuality of Engagement, inbound, for the PCA community because Mike is happy to make connections with the students and refers to these experiences as being "cute" and "fun" (also shown in Table 4 as the inbound example). By interaction with these two boys, Mike is able to feel valued by Pedro and Luis as a community member and is able to create meaningful changes in their identities.

Mike's involvement with PCA was furthered by the relationships that he built both with his peers (other physics students) and with the youth who were members of the community. Not only did these relationships seem to increase his enjoyment of PCA, but they also created a path for membership and allowed Mike to move inward toward a more central membership role. These connections took place on the way to the schools, during the activities at the schools, and even outside of the school setting (back at Mike's lab where he led a tour for the students) which allowed for his membership in PCA to overlap with his membership in the physics community.

### Negotiation of the Repertoire

By building relationships with the different members of the PCA community, Mike learned new skills and practices as well as developed a deeper understanding of the community norms. He talks about his interactions with the youth and how these interactions intersect with both PCA and physics community norms:

> I guess it seemed to me like the older brother who was maybe a little more, you know, attentive and observant, just kind of more of an observer, he would do well because of that, in science, and I think he would enjoy it. But the younger brother was also very, very skilled- like, he tended to be the one who would jump in and do something first. So, I think both of them would make good scientists. I think based on the personalities maybe it seems more likely that the older brother would go into science and the younger brother would just choose to do something else.

This reflection about some of the students Mike worked with was coded as Negotiation of the Repertoire for an insider member of the physics community because Mike is recognizing the strengths that allow one to succeed within physics (or science more broadly) within the two students. These interactions with the students impact his integration and therefore membership in the PCA community, as well as his desire to continue participating in the PCA because he sees these interactions as being in line with her personal desires to help others pursue science (as seen in his Accountability to the Enterprise example). He even goes on to say, "we're very happy to get the older brother. 'We', the scientific community, the dark side, are trying to recruit them. [laughter] But the younger brother is the kind of person we're trying to get through to it seems like. Clearly, he has the aptitude, and maybe if we can show him that then he would be more interested in doing it" (Accountability to the Enterprise for physics within the first sentence and Negotiation of the Repertoire of an insider member of physics for the second sentence).

For Mike, the intersection of physics and PCA takes place when both communities provide him with opportunities to engage in activities that align with his values and goal -- engage more people in physics-- while sharing his passion for physics. The PCA community is providing Mike a space to engage with practices that are important to him and to his ideas of being a physicist. But all of these practices have been, in some form or another, supported/fostered by the relationships he has formed with members of the communities or while engaging with the domain of the community.

## Intersecting Interests Codes

When we started coding the interviews with the CoP framework there was a need to specify what community the subjects were referring to. In some cases, while the experience happened in a specific community, the participants expressed that the experience also affected their membership in another community or area of interest. For example, when Cecilia discusses her hesitations for participating in PCA, she says that her hesitations were not related to PCA itself, but that time spent in PCA was time not spent in her research lab. In Figure 4, we show the intersection between PCA and other interests that were present in both Cecilia and Mike's interviews. The overlap between the areas of interest are represented in terms of the total amount of overlap that we saw with PCA. While we are representing the overlap between two areas of interest, it was possible for more than two interests to overlap.

Through the comparisons seen in Figure 4, we notice that there is no particular community that Cecilia or Mike perceive as having more connections with PCA; instead it seems as though participation in all these interests is balanced. Both seem to understand how the different interests and communities can intersect, allowing them to express their identity and therefore membership between their different interests and memberships without hindering participation in other communities or areas of interest. In fact, participation in PCA and the physics community compliments their values and motivation for participation in all of their interests. This confirms what we observed in the community dimensions, in which their *Accountability to the Enterprise* is connected to their passion for physics.

In particular for Cecilia, her connection with the other members of PCA sparked motivation not only within her physics research but also with her belief that continued participation in PCA was actually having an impact on the children and allowing her to feel like she was doing something significant. Cecilia's balance across the intersection of the interests can be linked to her path into physics and informal. During the interview, there are different occasions where she expresses that participating in informal physics is important because it allows her to provide those experiences to others that she did not have as a child, which - she believes - could have had a positive impact on her path to becoming a physicist. Cecilia also found that participating in informal programs became a source or motivation throughout her struggles in graduate school. Mike's case is driven by his desire to inspire others to feel the same passion he feels for physics.

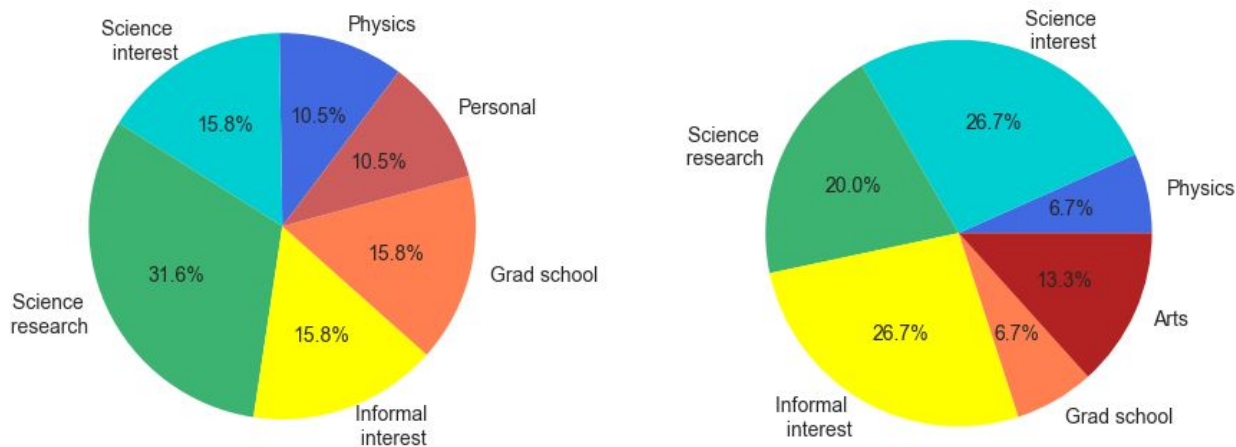

**Figure 4:** Percentage of overlap between PCA and other communities (out of total percentage of overlapping areas for PCA) for Cecilia (left) and Mike (right).

## Summary of Community Membership

For both, Mike and Cecilia, their membership to PCA, and the informal community more broadly, is linked to their physics identity. As physicists, they both believe, to varying degrees, that it is important to engage with the public to show them that physics can be awesome and accessible and that anyone can pursue their interest in physics. Therefore, we see the PCA domain aligning with their personal values and establishing a clear path to membership. Furthermore, participation in PCA also supports their physics identity by allowing them, as members of the physics community, to share their passion for the field (connected to their *Accountability to the Enterprise* in physics), demonstrate their knowledge of the content and practices of the field (*Negotiation of the Repertoire*), and engage with the children in physics practices (*Mutuality of Engagement*).

In Figure 5, we present the percentage of Community Dimensions for the PCA community seen within Mike's and Cecilia's interviews. The counts are normalized to the total number of codes; that is, the total number of codes in the interview would be treated as 100%. We chose to present the distribution of community dimensions for only the PCA community in order to paint a clearer picture of how their informal experience impacted other areas of interest. It is important to state that the length of the interview did not necessarily imply a higher number of codes overall (Mike's codes p/min = 4.64 and Cecilia's codes p/min = 6.22). In Figure 5, we are able to visually appreciate the differences in experience Mike and Cecilia had and how those are reflected through the community dimensions.

From Cecilia's community dimensions (see Figure 5 for details), we can infer that she perceives herself as an inbound member of the PCA community. She has a very good understanding of the domain of the community and continually contributes to this domain, which can be seen through her *Accountability to the Enterprise* (AE). However, she seems to still be developing the community (*Mutuality of*

*Engagement;* ME) and practice (*Negotiation of the Repertoire;* NR) dimensions. While Cecilia has definitely engaged in interactions with other members of the community, mostly the children that participate in the program (ME) as they work together on activities, and she has learned about the activities in PCA (NR), there are fewer number of codes in those dimensions, possibly due to the fact that she has only participated in PCA for one semester at the time of the interview and had to missed a couple of weeks during the semester.

Similarly, from Mike's distribution, we can infer that he mostly identifies as an inbound member of the PCA community. He has a very good grasp of the *Accountability to the Enterprise* dimension, which can be demonstrated by frequency of codes (see details in Figure 5). However, similar to Cecilia, Mike is still working on developing the practices and connections with other members of the community in order to become more a central member of the PCA community.

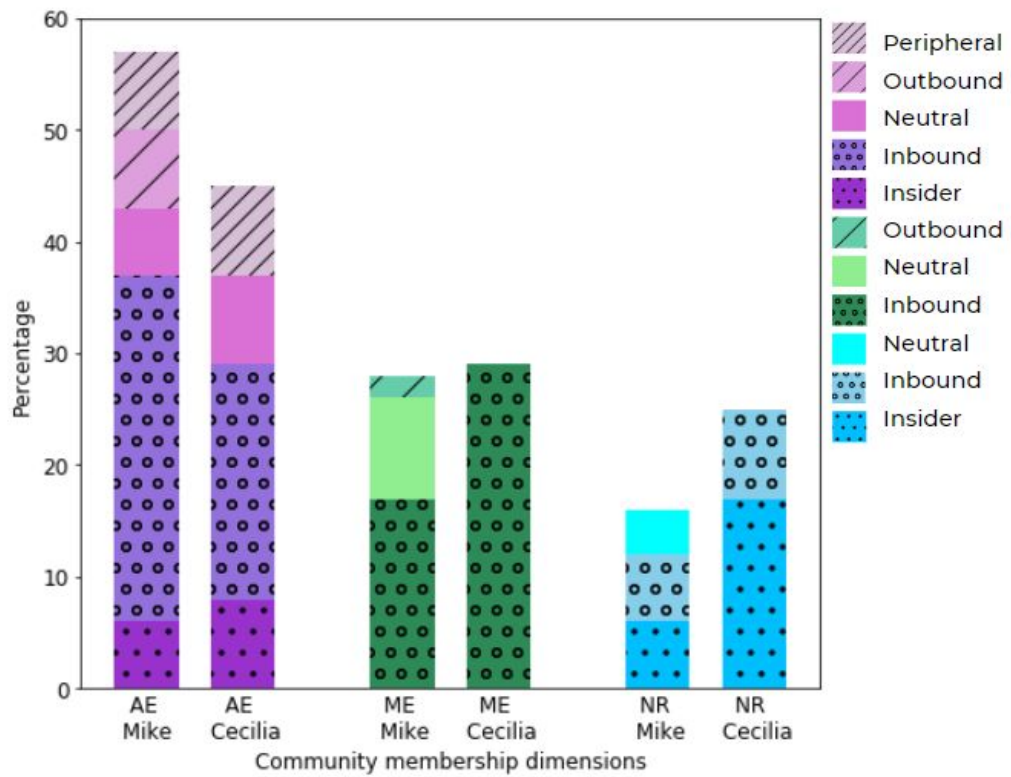

**Figure 5**: Community Dimensions within PCA represented as a percentage of all community membership codes for both Mike and Cecilia. The codes have been separated into the subcode categories - *Insider* (dots)*, Inbound* (circles)*, Neutral, Outbound* (diagonal)*,* and *Peripheral* (closely spaced diagonal) - to show the percentage of codes that were coded as movement or community membership levels. Mutuality of engagement did not produce any peripheral or insider codes and Negotiation of the Repertoire did not produce any outbound or peripheral codes.

From this case study, which only represents a snapshot of Cecilia and Mike's path towards membership within the PCA and the physics community, we notice that membership development is not a straight path but rather a gradual process. Developing membership, and therefore identity, is a dynamic process. We can move in and out of the community at different points of our membership - becoming more central members or retrieving to peripheral - depending on how we align with the community's dimensions and the experiences we have as members. At the moment of the interviews, both Mike and Cecilia had opened a path to become central members of the PCA community, however their membership levels in the three domains are dispersed, indicating that they are still exploring their positioning within PCA. While it is not required to be an insider member for all the domains to be considered a core member of the community, it is expected to see more cohesion.

## Mechanisms of Identity

Through the community dimensions we were able to characterize Mike's and Cecilia's membership within the PCA CoP. We noticed how membership shifts and changes based on how the community dimensions are encouraged/supported (or not) through their experiences in the program. These movements within levels of membership are often prompted by different mechanisms that impact the community membership dimensions. In this section, we show how the operationalized framework is able to capture what identity mechanisms have contributed to Cecilia and Mike's membership within the PCA and physics community. In Figure 6, we present the percentage of Mike and Cecilia's identity mechanisms that were connected to their participation in the PCA community. As with the community dimensions, the counts are normalized to the total number of codes present in the individual interviews and then we calculate how much of that represents each of the mechanisms. Here we will discuss the mechanism and what it meant for Cecilia and Mike during their participation in PCA.

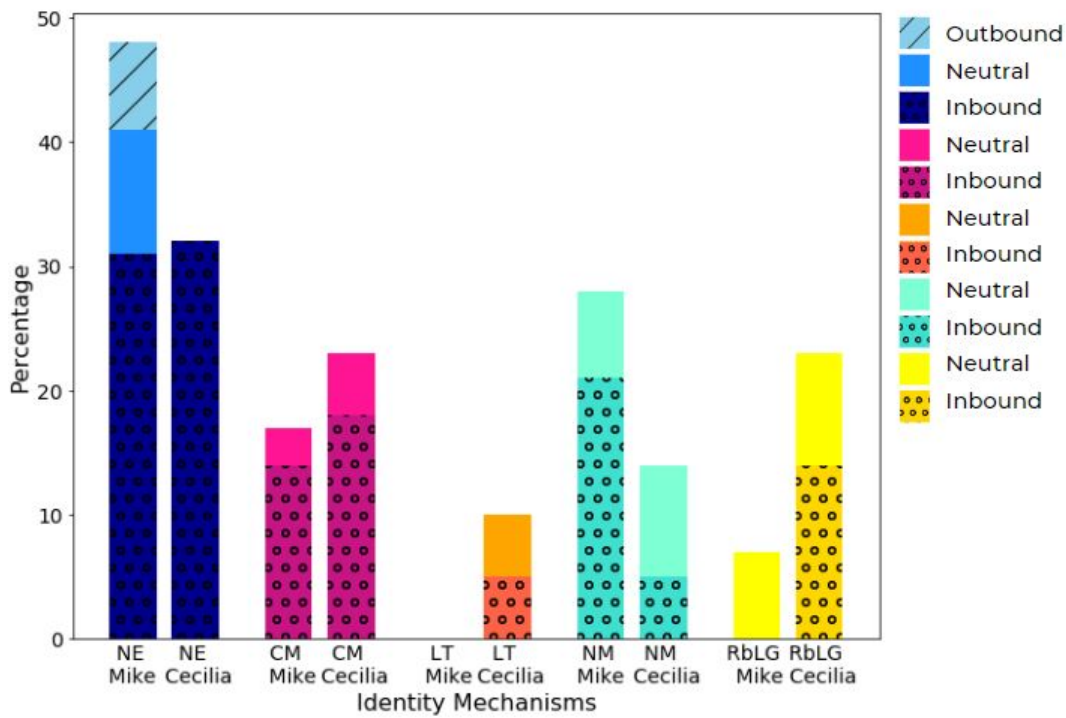

**Figure 6**: Identity mechanisms for PCS shown as a percentage of all identity mechanism codes for both Mike and Cecilia. The codes have been seperated into the subcode categories - *Inbound* (circle), *Neutral*, and *Outbound* (diagonal lines) - to show movement between membership levels. Negotiated experiences (NE) was the only category with outbound codes.

# Mike

## Negotiated Experiences

Through Mike's story, and the codes reflected in Figure 6, we can see that the most influential mechanism for Mike's movement within the PCA community is from interactions with members of the community and how these interactions bring meaning to his participation. Mike's *negotiated experiences* with others in the PCA community of practice allow him to connect to a sense of belonging in the community and to be recognized as a member. These experiences show how the interactions with the other members, particularly the children, reaffirm Mike's motivation for participation and to contribute to the domain of the PCA community.

Mike's motivation for participation in the PCA community is also linked to his participation in the physics community. This can be seen through his comments about what prompted him to continue participating in PCA and what he saw the students getting out of their experiences:
> I think really it just came down to the fact that they [the children] were getting involved in [the activities], like they were participating, you know? That right there is the statement I really looked for. I don't know if any of them are more likely to become scientists, but you kind of imagine so; or at least more likely to think that it's a possibility. Whether they'll actually want to do that or not, you know, at that point in their life.

The above quote from Mike's interview was coded as Negotiated Experiences for an inbound trajectory in the PCA community because he is talking about the positive interactions with the children and that he believes they want to become scientists. Also, when Mike was asked about his favorite moment while participating in PCA, Mike recalls, "That was when Pedro asked if he could work in the lab, so that was my favorite" (inbound trajectory in PCA for Negotiated Experiences). Both of these quotes reflect how Mike's membership in PCA allows him to engage more people in physics while simultaneously sharing his passion for physics. By having positive interactions with the children, and seeing them truly engage and enjoy the activities, his membership within the PCA and physics communities are reaffirmed. Furthermore, Mike mentions that he has formed brotherly connections to some of the children in the program, which indicates a strong sense of belonging to the community. While this connection was during that particular semester, it could still be a motivation for continuous participation because he is wanting to create more connections. Mike also mentions that interactions with his peers - physics graduate students that facilitated PCA activities - were positive, always fun and stress relievers. These interactions may have also contributed to his sense of belonging in both communities.

## Community Membership

Beyond those connections, Mike also reflects on the forms of competence and skills he learned through his participation in PCA:
> I think I definitely have learned a lot. And this is something I wanted to get out of [PCA] too; I knew I would learn a lot about what it takes to teach, especially young kids, science, and expose them to ideas. I think I definitely developed- What I've gotten out of it is just being a better teacher, that's for sure, having a better understanding of- but also a better sense of what schools are like these days and what kids are like these days; you know, what it takes to get them into science.

Mike's reflection was coded as *Community Membership-inbound* for PCA, because he is talking about the teaching capabilities that he developed as a member of PCA. Mike's reflection on how to teach the youth about science and to co-think about experiments with them shows that these experiences were an essential mechanism in terms of him becoming a central member of PCA. Furthermore, Mike comments on interactions that he experienced with the classroom's teachers while in PCA, "[H]e (the science teacher) was really helpful in- it was really informative to me to see how he took what I said and explained it to them. I was trying to make it accessible, but he really knew how to do that, so that was cool." When Mike was talking about how the science teacher explained things to the students, he was sharing how another member of the PCA community helped him learn how to be a better member in PCA, therefore coded as *Community Membership-inbound* for PCA. These instances indicate that Mike felt he was growing as a member of the PCA community and moving towards a central member by understanding and learning the practices of the community through interactions. This sense of growth and competence are important factors for continuous participation in a community of practice and for building identity.

Nexus of Multimembership

The most important role as a member of PCA for Mike, is to convey to children the confidence or push they need to believe that they can do science. He reflects on how an external push helped him pursue physics as a career. Therefore, he is now providing that push to the youth in PCA: "To get more people into science- because I think certainly that's obviously the goal of PCA, or at least to encourage kids and show them that they can do it if they wanted. And that's a big thing, is motivation and self-esteem stuff." This reflection from Mike was coded as Nexus of Multimembership (neutral) for PCA and science interest because it is showing how Mike connects the goal of PCA and with being interested in science but he is not speaking about this in a positive or negative sort of way. This goal of PCA aligns with Mike's own values and is possibly a reason for his commitment to the program. His constant negotiation for participation between physics and PCA appears to have an impact on Mike's membership movement within PCA, which was reflected in the community dimension of *Accountability to the Enterprise* (see Table 1 for the inbound example).

Mike's personal goals and mission aligned with PCA domain, which is intertwined with his participation in physics; as he becomes a more central member in the physics CoP, it is important to him to share his passion for physics, engage people in physics, and help them build the confidence necessary to pursue a career in science. Being involved in PCA has allowed Mike to do all of those things that are important in the physics community. He comments, "But yeah, kind of the more I've grown as a scientist, the more I've wanted to help others get into science." This reflection about wanting to share his experiences in science with others was coded as Nexus of Multimembership with an inbound trajectory in both the informal community and science research because Mike is explaining how his growth as a scientist influences his growth in the informal community. This connection permeates in the codes when he discusses his constant negotiation between physics and PCA. However, while Mike believes that graduate school was a good time to start doing more outreach, he is also often conflicted between sharing his time between graduate school and outreach programs. He expresses this by saying, "Yeah. I wouldn't've started [PCA] if I were partway into the first semester, you know, and I knew how intense it would be [referring to the semester]. But yeah, [PCA] is not a big time commitment […] And even when the concept of going [to PCA] was stressful, going itself was very cathartic and nice. It was always fun, as it was this semester. It was always just a stress reliever" (Nexus of Multimembership, neutral, for both graduate school and PCA).

Overall, Mike's experience reflects that his growth as a member of the physics community leads to a membership in PCA and informal physics community more broadly, in order to achieve his goal of spreading a passion for science and physics. In a similar way, the more he participates and becomes a more central member of the PCA community his membership in the physics community gets reaffirmed. Therefore, the Nexus of Multimembership between the physics and PCA communities for Mike can be seen as a mutualistic symbiotic relationship in which his participation in one strengthens his participation in the other and vice versa.

## Cecilia

Unlike Mike, whose code frequency was more heavily concentrated on some of the identity mechanisms, Cecilia's movement in PCA seems to be driven by all the identity mechanisms. In Figure 6, we see that Cecilia's experiences are represented throughout all five of the mechanisms.

### Negotiated Experiences

As it was for Mike, Cecilia's connections with the children in the program seemed to have an impact on her membership. For Cecilia, the interaction with the children helped reaffirm her motivation for participation in PCA; seeing the children positively respond to engaging with the physics activities made her feel as though the goal was being accomplished. When asked about her interactions with the students, she comments, "I was surprised at how enthusiastic all the kids were, and that was something that definitely took me by surprise. I guess it's because it's a voluntary thing." This quote in Cecilia's interview was coded as inbound Negotiated Experiences for the PCA community because she is excited by her interactions with the children. The enthusiasm displayed by the youth participants gave Cecilia a sense of belonging because she was recognized and able to build up a rapport with them. She shares this when she comments on how the children recognized her over time as she kept coming back to the schools and one student even told her that was their favorite (see the inbound example in Table 2).

However, these experiences also impact Cecilia's membership within the physics community. She is not only able to share her passion for physics and receive positive responses from the children, but the children also encouraged her continuous participation in the physics community. There was an instance (detailed in Cecilia's Mutuality of Engagement section) where one of the students participating in PCA commented on Cecilia having "the coolest job ever" (which was also coded as an inbound Negotiated Experience for graduate school). This reaction to being a graduate student was something that encouraged Cecilia to take that viewpoint about her job, even when the going is tough.

In addition to talking about how she was able to interact with the youth in the classroom space during PCA, Cecilia mentions experiences where the youth come into her space within the physics community. As part of PCA, the students are able to go on lab tours at the university and Cecilia was able to show her lab and research to the PCA children:
> So it's really cool, and that's always one of my favorite parts is showing the kids around the lab. I just, I love that because they ask all these questions and I just eat it up. So I don't know, it's been really neat for kids to be like 'wow, that's really cool!' It made me feel like I'm actually doing something.

This remark about the lab tours was coded as *Negotiated Experience-inbound* for both PCA and science research because Cecilia loved interacting with the children about her own work. These experiences, both during the time spent in classrooms after school and in the labs on campus, allowed for Cecilia to move toward a more central membership role within PCA and the physics community because she is able to have positive experiences with the children that allow her to appreciate her role in both communities.

### Learning Trajectory

Cecilia's childhood experiences (or lack of) with informal programs seems to be a driving force for her continuous participation in PCA and informal programs more broadly. She shares that when looking back, she wishes there had been more access to opportunities that would allow her to engage with science and physics sooner. We can see these lack of experiences impacting how Cecilia views outreach now when she comments on how she wishes someone had introduced her to science at a younger age and now she believes it is very important to educate others (see Cecilia's Accountability to the Enterprise section for the quote). This instance of Cecilia recognizing the importance of educating youth was also coded as an inbound Learning Trajectory for both the informal/outreach community and the science interest community because she is relating her past and present experiences to her involvement in both science and PCA.

However, Cecilia also takes the opportunity to reflect on ways that she was encouraged to pursue STEM as an option. She believes that part of her continuation in participating in physics has been because of positive role models and figures throughout her career that have instilled her with confidence and the idea that she is capable of doing physics. She shares how influential some teachers have been when she says, "My high school chemistry and my high school physics teacher, I had them both actually again my senior year because I took the AP level of both of those, and they were just like yes, you can do this, this is a thing that you can do for the rest of your life. And that was just, it had never even dawned on me" (inbound Learning Trajectory for physics and science interest). She knows that these experiences were impactful in her own trajectory through science and so she speaks about offering similar experiences to the student participants in PCA.

Furthermore, outreach became an important part of her participation in the physics community and that positive experience lead her to seek other similar opportunities:
> [I] was actually a grad student for like a month and a half in [abroad physics institution]... [S]hort story long, I ended up dropping out, but for that month and a half, couple of months that I was there, outreach was really big[...] [T]here were multiple times that I went to the university and helped with outreach for kids or, you know, went to like a mall in the city and helped with outreach for kids[…] when I came back to [U.S university] I wanted to get back into that in some situation. So, when I finally had the chance and the guts to finally say okay, I'm going to do this once a week.

In this quote Cecilia is expressing how past experiences have been key to determine her participation in the different communities and thus it was coded as *Learning Trajectory* (outbound for grad school and inbound for informal interest). Her negative experience in the abroad physics institution where she did a semester had a positive side which was participation in informal physics activities. This then led her to want to pursue more of those positive experiences with informal physics. Therefore, experiences categorized under *Learning Trajectory* show that these experiences were a very important mechanism for Cecilia's participation in both the PCA and the physics community because much of her past and present revolved around being interested in science and being interested in informal/outreach opportunities.

## Community Membership

For Cecilia, engaging in the practices of PCA helped her gain new perspectives and ways of looking at the world. When asked what she has gained from PCA, she comments on how she has gained an appreciation for the spectrum of people who can participate in physics, and science in general (see Negotiation of the Repertoire section for Cecilia's quote). This instance in her interview was also coded as a Community Membership on an inbound trajectory for PCA and science interest because Cecilia is relating her experiences in PCA with her views on science as a community. Furthermore, participation within the PCA community also strengthens her connection with the practices of the physics community because the children recognize her as an expert of the physics domain. This recognition allows Cecilia to feel more competent in the physics practices and she share this when reflecting on what she has gained from PCA, "the biggest benefit that I've gotten is, you know, I don't want to say the ego boost, but kind of you know having kids come up to you and be like 'wow, this is really cool!" (inbound Community Membership for PCA). Therefore, by being a member of the PCA community, Cecilia is not only learning new skills, practices, and norms of interactions, but through the connections with members of the PCA community she is also gaining appreciation for her membership in the physics community.

## Nexus of Multimembership

As with Mike, Cecilia's negotiation of participation in the physics community and PCA has had an impact on her membership within both communities. She also has struggled to manage the time spent on both communities, but she was more driven by pressure from the physics community to not engage in activities outside her own research. She shares this as a reason for questioning if she should get involved with PCA when she says:

My only real hesitations weren't related to PCA necessarily, they were more related to, you know, being able to escape the lab. And that's really just a laboratory politics thing, that really doesn't have much to do with PCA […] You know, the time lost in the lab I could always make up later. And it was really just, it was just fun. Like I might do a

> semester off semester on sort of thing, but really that would be more to give other people a chance to experience it. I enjoyed PCA and I hope it continues.

This sense of hesitation and her reasoning for being involved with PCA was coded as a neutral *Nexus of Multimembership* experience for both PCA and the science research community because Cecilia is talking how PCA and her lab work are at odds with each other. In Cecilia's case, membership in the PCA community seems to clash with her membership in the physics community mainly due to recognition by members of the physics community and the lack of support that she receives from others in her research lab. However, she sees outreach as an important element of being a scientist and plans to continue on it throughout her career path. This is clear when asked whether she can see herself participating in physics without doing outreach and she responds by saying, "It would be kind of lame. I mean I could, and I did for the first four years I was in grad school, or first three years I was in grad school. But I always missed it when I wasn't a part of it. You know, I don't think it's essential for me, you know, to get research done [...] but I kind of have this obligation to help educate that I've felt" (inbound Nexus of Multimembership for physics and the informal community). Through these thoughts from Cecilia, we can see that she struggles with being involved with both outreach and physics, but she persists because the goal of outreach is personally very important to her.

Relationship between Local and Global

Cecilia feels a deep obligation to educate people and engage them with science, and physics in particular. This obligation is driven by her sense of belonging to the physics community and wanting to contribute to engaging more people in that community. She has expressed her sense of obligation in different forms throughout her participation in the PCA community and how her participation in that community connects to a bigger purpose. She shares:

> You know, it was more kind of like how can I help people, you know. You hear about people like, you know, I developed an organization that builds, you know, bathrooms in Africa for underprivileged people that don't have bathrooms, and I'm like wow, that's cool. But I'm just, I'm not built that way. Kind of the way that I've always thought that I could make the most impact in this world is to try to advance science knowledge, and so that's kind of at the end of the day what gets me through. The shorter answer though is that it's awesome.

This reflection on how Cecilia's participation in PCA can be related to the world more broadly, was coded as *Relationship between Local and Global-inbound* for both PCA and the physics community. This is because she is recognizing that her strength in this world is related to telling others about science. She sees the connection between participating in PCA, the local impact (getting the group of children that participate more interested in physics/science) and how that translates to a bigger picture (changing society's perception of physics/science). Cecilia's perception of how the local impact connects to a larger picture becomes a big driver for her membership in the PCA community, the outreach community, and the physics community more broadly. This connection can be seen when she shares her thoughts on who gets to do science and why the broader community should be involved:

> And so maybe in a sense that's what I get out of it, is that when people say science is awesome and we should pursue it, they eventually go vote and, you know, the NSF gets a bunch of money and I get to take some of that money and go do science with it in that sense. But it's kind of a more long-term thing. Like I just want people to, I want everybody to understand that like science isn't out to get you, it's not going to, you know, attack your belief system unless you let it. You know, it's here to kind of save us from ourselves in a sense. (inbound Relationship between Local and Global)

Overall, Cecilia shares a variety of experiences and views that impact how she participates within PCA and the physics community. These experiences are both from her past (experiences from her childhood) and through her present participation in both communities. During the interview, Cecilia shares how interacting with other members of the PCA community and how experiences within PCA have impacted her own perception of her membership within physics. She also shares how these experiences and interactions engage with her previous trajectories and her worldview.

## Summary of Mechanism Constructs

In this section, we examined what mechanisms of identity within the operationalized CoP framework impacted Mike and Cecilia's membership in the PCA community. Additionally, we examined how these mechanisms linked participation in PCA with participation in the physics community. As noted in the community dimensions section, Mike and Cecilia's membership within PCA is connected to their value of engaging with the public, particularly children, in order to show audiences that physics can be awesome and anyone can pursue it. Therefore, we expected to find that the mechanisms that had a larger impact on their membership were related to how their participation supported their goal of engaging more children in physics.

In Figure 6, we notice that for both Mike and Cecilia, their relationships with other members (*Negotiated Experiences*) are the most important mechanism within the PCA Community of Practice to move them towards building identity. This is in agreement with other frameworks that have studied identity and sense of belonging, in which being recognized by others in the field was essential to form identity [1-2,28,55].

In Mike's interview, we can see that *Negotiated Experiences* (about 50% of his overall mechanism codes, seen in Figure 6) was appreciated through his comments about the brotherly connection developed with some of the children in the program and, more specifically, that these interactions allowed him to achieve his goal of seeing children positively interact with physics. He shares that the youth participants even hinted at working with him in the lab when they grow up or becoming scientists. The second most relevant mechanism present in Mike's path to membership is the *Nexus of Multimembership* mechanism (in figure 6, we see that this is about 30% of all of the mechanism codes). In this case, it is Mike's strong sense of belonging within the physics community and his desire to share that with the children so they can also pursue a career in science that inspires him to continue participating in PCA. This desire to share his physics experiences with the youth participants was especially true after experiencing positive interactions with the children during PCA. Finally, the third relevant mechanism that appeared in Mike's coding was connected to how his participation in PCA allowed him to learn more about communicating and engaging the children in science (*Community Membership).*

For Cecilia, we observed that all the mechanisms of the framework contributed in some form or another to her membership in the PCA community (in fact, we only saw Learning Trajectory in Cecilia's interview). Like Mike, the connection and interactions with the children in PCA and their positive responses to engaging in physics activity was definitely the most significant factor that influenced her membership. Her main driver for participating in the PCA community is her personal obligation to educate others, children in particular, about how exciting physics is and the purpose of physics. In her case she seems to perceive that all the mechanisms contribute to achieving her goal.

Finally, it is interesting to note that even though the structures of the program and the degree of participation for Mike and Cecilia are similar (going to the school once a week to facilitate some activities), their experiences were different enough that Cecilia perceived support of all the identity mechanisms, while Mike only discussed a subset of the mechanisms. For future work, we are interested in studying whether support from all the mechanisms is important or even necessary for becoming a central member of the community or if individuals can become central members by only experiencing some of the mechanisms.

## Implications & Future work

We started this paper asking whether the CoP framework was a powerful instrument to understand identity formation within informal programs and how student facilitators view their membership within the program. This use of the framework has allowed us to study identity formation from a sociocultural perspective. Furthermore, it can help us identify structures within informal physics programs that could be changed to better support identity formation for students, particularly those from underrepresented groups in physics. Consequently, the CoP framework can be used as a tool to make the field of physics more inclusive. We see that the operationalized framework allows us to understand how participation in informal physics programs may impact university physics students' identity and sense of belonging within the physics field. The power of this framework is that it allows us to sense where the university students see themselves within the Community of Practice and to learn what aspects of that community impact their involvement. From the point of view of physics identity, this is important because we can understand how participation in informal programs allows for the growth of a physics identity and how these students can become more central members within the physics community. In this framework, our diverse backgrounds and multiple identities represent a key mechanism for identity development, because it is only when our different memberships are able to intersect, collaborate, and work together that we can fully become integral members of a community of practice.

The coded narrative of Mike and Cecilia allows us to understand the connection between the community dimensions and the mechanisms of identity. This connection between these constructs within the framework allow us to understand how facilitators view their membership within the informal programs and what specific program structures influence identity development. First, we used the Community Dimensions to establish an understanding of Cecilia and Mike's membership in PCA and how they perceived their own positioning within it. Figure 7 shows the identified recurring themes across all three community dimensions for both Mike and Cecilia. While both interviewees show commitment and work towards the domain of the PCA community, Cecilia's connection to the domain seems to be set within a broader context. She talks about the domain from the perspective of outreach in general rather than just from the point of PCA and sees this as her duty as a physicist to help change the narrative of who can be involved. However, in Mike's case, he sees his participation in PCA as a way for him to share his knowledge and passion for physics in a new way. Both of them see their desire to engage more children in physics connected with their participation in both PCA and physics. By picking out the recurring themes among all three community dimensions, we are able to understand what elements of the PCA community that Mike and Cecilia relate to such that they want to participate in the program. An understanding of these themes can influence how future informal physics programs are structured because programs should be aligned to the interests of student facilitators in order to be an impactful experience for those facilitators.

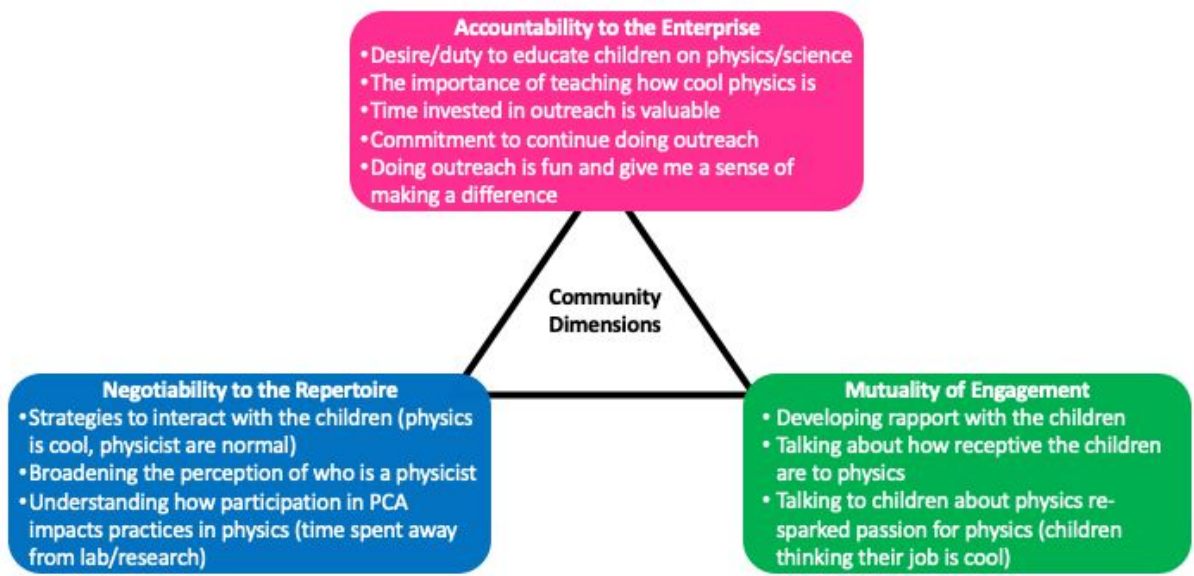

**Figure 7**: Themes within the community dimension codes for Mike and Cecilia. The themes are emergent from the codes that were identified within each of the three dimensions.

As with the community dimensions, some recurring themes were identified for each mechanism. These themes can point out what particular elements or structures provided by the community (and therefore the design of the program) can (or not) help foster individual identities. Figure 8 contains the themes prevalent within the coding for each of the identity mechanisms. We see that Mike and Cecilia both found the enthusiastic responses of the children as an integral part of their participation and contributing to the feeling that they were making a difference and providing role models to others. Additionally, both Mike and Cecilia seemed to be constantly negotiating between their membership in the physics community and the PCA community. In part, it is their membership in physics that drives them to engage in PCA because they want to share their passion but some of the norms and practices of physics conflict with their participation in PCA. At the same time, Mike and Cecilia's participation in PCA seems to be shifting their perception of what it means to be a member of the physics community by broadening their personal definition of what constitutes the physics community practices and who can be in it. Being able to understand what mechanisms influence identity within the community is important for understanding how informal physics programs can support physics students who act as facilitators. When researchers and practitioners have a better understanding of the specific elements of informal programs that are influential, then we can reform and better design future programs.

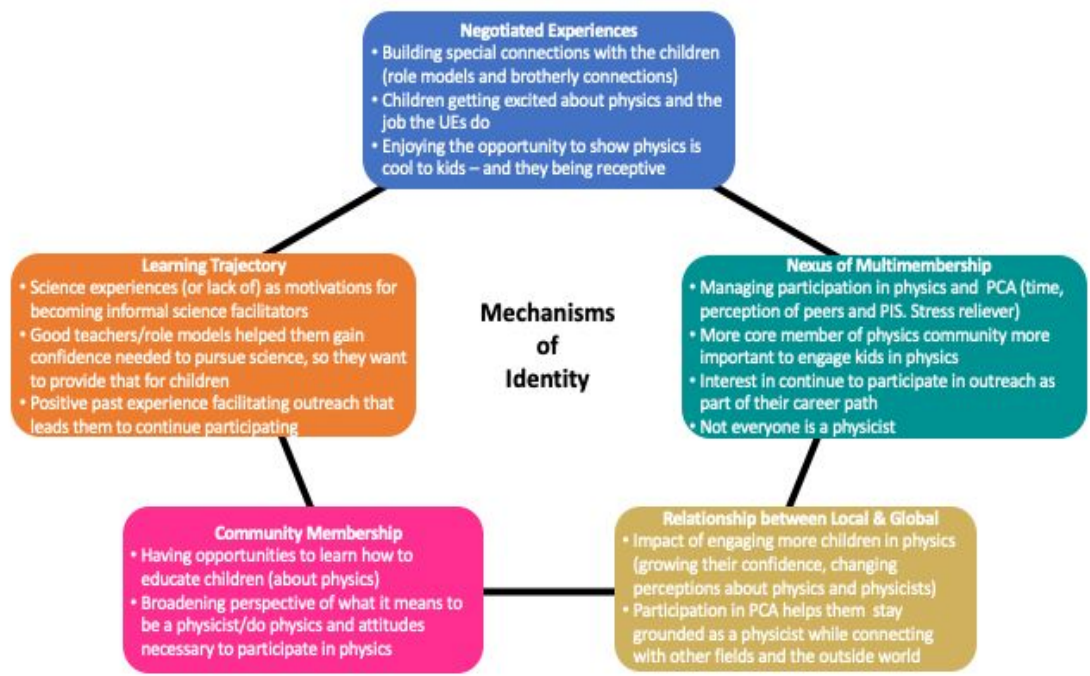

**Figure 8**: Themes within the identity mechanisms codes for Mike and Cecilia. The themes are emergent from the codes on the corresponding mechanisms.

Therefore if we revisit our research question on whether facilitating informal physics programs can help foster university students' identity, we can use the community dimensions and mechanisms of identity from Mike and Cecilia to understand how the operationalized framework illuminates positive identity development. The framework helped us understand how Mike and Cecilia perceived their membership within PCA and physics as well as what structural elements of PCA aided in membership, or identity, growth. Additionally, when considering diversity and inclusion in physics, this operationalized CoP framework could be utilized to understand how specific programs support identity development. We believe that the consequences of this framework reach far beyond the scope of informal environments and that it could be used to understand how physics, as a field, could be inclusive and welcoming to multiple different identities and experiences.

To continue this work, we plan to analyze the remaining interviews in our data set as well as use interviews from another program to look at facilitators from a variety of backgrounds, disciplines, and levels within those disciplines, as well as different models of informal programs. We will use the collection of many individual experiences to understand the programmatic experience of facilitators in each informal physics program. The purpose of this would be to understand what structures within informal programs have a larger influence on discipline-based identity development and how these structures can change based on program design. Implications of identifying what structures or experiences help reinforce or foster university students' many different identities, both physics and informal engagement, can lead to designs of more inclusive environments not only within informal environments but also in formal learning spaces. By creating spaces in which students feel identified, we increase the possibility for more students from diverse backgrounds to participate and persist in the field.

## Acknowledgements

We would first like to acknowledge and thank our study participants, who have shared their time, energy, and stories to contribute to this work. Additionally, this paper is based upon work supported by the European Research Council through Marie Skłodowska-Curie Actions Individual Fellowships (MSCA-IF) and NSF Advancing Informal STEM Learning award 1423496. Any opinions, findings, and conclusions or recommendations expressed in this material are those of the author(s) and do not necessarily reflect the views of the European Research Council or the National Science Foundation.

## References

[1] "Bachelor's Degrees Earned by Women, by Major" *American Institute of Physics: Education and Diversity.* Retrieved September 9, 2019 from
https://www.aps.org/programs/education/statistics/womenmajors.cfm.

[2] "Physics Degrees Earned by Underrepresented Minorities" *American Institute of Physics: Education and Diversity.* Retrieved September 9, 2019 from
https://www.aps.org/programs/education/statistics/phdpopulation.cfm.

[3] Rosa, K. (2019). Race, Gender, and Sexual Minorities in Physics: Hashtag Activism in Brazil. In Upgrading Physics Education to Meet the Needs of Society (pp. 221-238). Springer, Cham.

[4] National Academies of Sciences, Engineering, and Medicine. (2018). Sexual harassment of women: climate, culture, and consequences in academic sciences, engineering, and medicine. National Academies Press.

[5] Rodriguez, I., Potvin, G., & Kramer, L. H. (2016). How gender and reformed introductory physics impacts student success in advanced physics courses and continuation in the physics major. Physical Review Physics Education Research, 12(2), 020118.

[6] Barthelemy, R. S., McCormick, M., & Henderson, C. (2016). Gender discrimination in physics and astronomy: Graduate student experiences of sexism and gender microaggressions. Physical Review Physics Education Research, 12(2), 020119.

[7] Hughes, B. E. (2018). Coming out in STEM: Factors affecting retention of sexual minority STEM students. Science Advances, 4(3), eaao6373.

[8] Eaton, A. A., Saunders, J. F., Jacobson, R. K., & West, K. (2019). How Gender and Race Stereotypes Impact the Advancement of Scholars in STEM: Professors' Biased Evaluations of Physics and Biology Post-Doctoral Candidates. Sex Roles, 1-15.

[9] Lewis, K. L., Stout, J. G., Pollock, S. J., Finkelstein, N. D., & Ito, T. A. (2016). Fitting in or opting out: A review of key social-psychological factors influencing a sense of belonging for women in physics. *Physical Review Physics Education Research*, *12*(2), 020110.

[10] Hausmann, L. R., Schofield, J. W., & Woods, R. L. (2007). Sense of belonging as a predictor of intentions to persist among African American and White first-year college students. Research in higher education, 48(7), 803-839.

[11] Rainey, K., Dancy, M., Mickelson, R., Stearns, E., & Moller, S. (2018). Race and gender differences in how sense of belonging influences decisions to major in STEM. International journal of STEM education, 5(1), 1

[12] McGee, E. O. (2016). Devalued Black and Latino racial identities: A by-product of STEM college culture?. *American Educational Research Journal*, *53*(6), 1626-1662.

[13] https://www.aps.org/meetings/policies/code-conduct.cfm

[14] https://www.aapt.org/aboutaapt/organization/code_of_conduct.cfm

[15] Corbo, J. C., Reinholz, D. L., Dancy, M. H., Deetz, S., & Finkelstein, N. (2016). Framework for transforming departmental culture to support educational innovation. Physical Review Physics Education Research, 12(1), 010113.

[16] Albanna, B. F., Corbo, J. C., Dounas-Frazer, D. R., Little, A., & Zaniewski, A. M. (2013, January). Building classroom and organizational structure around positive cultural values. In AIP Conference Proceedings (Vol. 1513, No. 1, pp. 7-10). AIP.

[17] Holmegaard, H. T., Madsen, L. M., & Ulriksen, L. (2014). To choose or not to choose science: Constructions of desirable identities among young people considering a STEM higher education programme. International Journal of Science Education, 36(2), 186-215.

[18] Z. Hazari, G. Sonnert, P. Sadler, and M.-C. Shanahan. (2010). Connecting high school physics experiences, outcome expectations, physics identity, and physics career choice: A gender study, J. Res. Sci. Teach. 47, 978.
44

[19] O. Pierrakos, T. K. Beam, J. Constantz, A. Johri and R. Anderson, "On the development of a professional identity: engineering persisters vs engineering switchers," *2009 39th IEEE Frontiers in Education Conference*, San Antonio, TX, 2009, pp. 1-6.
doi: 10.1109/FIE.2009.5350571

[20] Chinn, P. W. (2002). Asian and Pacific Islander women scientists and engineers: A narrative exploration of model minority, gender, and racial stereotypes. Journal of Research in Science Teaching: The Official Journal of the National Association for Research in Science Teaching, 39(4), 302-323.

[21] Hazari, Z., Sadler, P. M., & Sonnert, G. (2013). The science identity of college students: Exploring the intersection of gender, race, and ethnicity. Journal of College Science Teaching, 42(5), 82-91.

[22] Barton, A. C., & Tan, E. (2010). We be burnin'! Agency, identity, and science learning. The Journal of the Learning Sciences, 19(2), 187-229.

[23] MacPhee, D., Farro, S., & Canetto, S. S. (2013). Academic self‐efficacy and performance of underrepresented STEM majors: Gender, ethnic, and social class patterns. Analyses of Social Issues and Public Policy, 13(1), 347-369.

[24] Cochran, G. L., Hodapp, T., & Brown, E. E. (2018, March). Identifying barriers to ethnic/racial minority students' participation in graduate physics. In Physics Education Research Conference.

[25] Hyater-Adams, S., Fracchiolla, C., Finkelstein, N., & Hinko, K. (2018). Critical look at physics identity: An operationalized framework for examining race and physics identity. *Physical Review Physics Education Research*, *14*(1), 010132..

[26 ] Hyater-Adams, S., Fracchiolla, C., Williams, T., Finkelstein, N., & Hinko, K. (2019). Deconstructing Black physics identity: Linking individual and social constructs using the critical physics identity framework. Physical Review Physics Education Research, 15(2), 020115.

[27] Meyers, K. L., Ohland, M. W., Pawley, A. L., Silliman, S. E., & Smith, K. A. (2012). Factors relating to engineering identity. Global Journal of Engineering Education, 14(1), 119-131.

[28] Hennessey, E., Cole, J., Shastri, P., Esquivel, J., Singh, C., Johnson, R., & Ghose, S. (2019, June). Workshop report: Intersecting identities—gender and intersectionality in physics. In AIP Conference Proceedings (Vol. 2109, No. 1, p. 040001). AIP Publishing.

[29] Williams, T. (2018). The Intersection of Identity and Performing Arts of Black Physicists (Dissertation).

[30] Auguste, D. A Mixed-methods Study of Non-text Social Media Content as a Window into African-American Youth STEM Identities. (2018). Presented in the American Society for Engineering Education Annual Conference, Salt Lake City, Utah. Paper ID #21942

[31] Close, E. W., Close, H. G., & Donnelly, D. (2013, January). Understanding the learning assistant experience with physics identity. In *AIP Conference Proceedings* (Vol. 1513, No. 1, pp. 106-109). AIP.

[32] Close, E. W., Conn, J., & Close, H. G. (2016). Becoming physics people: Development of integrated physics identity through the Learning Assistant experience. *Physical Review Physics Education Research*, *12*(1), 010109.

[33] Irving, P. W. and Sayre E. C. (2015). Becoming a physicist: The roles of research, mindsets, and milestones in upper-division student perceptions. *Physical Review Physics Education Research, 11(2), 020120.*

[34] Stein, F. M. (2001). Re-preparing the secondary physics teacher. *Physics Education*, *36*(1), 52.

[35] Eylon, B. S., & Bagno, E. (2006). design model for professional development of teachers: Designing lessons with physics education research. *Physical Review Special Topics-Physics Education Research*, *2*(2), 020106.

[36] J. Lave. (1991). Situating learning in communities of practice, in Perspectives on Socially Shared Cognition, edited by L. Resnick, J. Levine, and S. Teasley (APA, Washington, DC), pp. 63–82.

[37] J. Lave and E. Wenger. (1991) Situated Learning: Legitimate Peripheral Participation (Cambridge University Press, Cambridge, England).

[38] Dierking, L. D. (2014, June). Cascading influences: Long-term impacts of informal STEM experiences for girls. In *27th Annual Visitor Studies Association Conference, Albuquerque, New Mexico*.

[39] Fracchiolla, C., Hyater-Adams, S., Finkelstein, N., & Hinko, K. (2016, July 20-21). University physics students' motivations and experiences in informal physics programs. Paper presented at Physics Education Research Conference 2016, Sacramento, CA. Retrieved September 23, 2019, from
https://www.compadre.org/Repository/document/ServeFile.cfm?ID=14215&DocID=4568

[40] Hinko, K. A., Madigan, P., Miller, E., & Finkelstein, N. D. (2016). Characterizing pedagogical practices of university physics students in informal learning environments. Physical Review Physics Education Research, 12(1), 010111.

[41] Prefontaine, B., Fracchiolla, C., Vasquez, M., & Hinko, K. (2018, August 1-2). Intense Outreach: Experiences Shifting University Students' Identities. Paper presented at Physics Education Research Conference 2018, Washington, DC. Retrieved September 23, 2019, from
https://www.compadre.org/Repository/document/ServeFile.cfm?ID=14841&DocID=4988

[42] Fracchiolla, C., Prefontaine, B., Vasquez, M., & Hinko, K. (2018, July 9-13). Is participation in Public Engagement an integral part of shaping physics students' identity? Paper presented at the GIREP-MPTL conference 2018, San Sebastian - Spain. (in Press)

[43] Mullen, Claire, Prefontaine, B., Hinko, K., and Fracchiolla, C. (2019, July 24-25). Why it should be "and" not "or": Physics and Music. Paper presented at Physics Education Research Conference 2019, Provo, UT. (in Press)

[44] Holland, D., Lachicotte, W., Skinner, D., & Cain, C. (1998). Agency and identity in cultural worlds. Cambridge, MA: Harvard University Press. Jarvis, J., & Robinson, M. (1997). Analysing Educational Discourse: An Exploratory Study of Teacher Response and Support to Pupils' Learning1. Applied Linguistics, 18(2), 212-228.

[45] Marcia, J. E. (1980) Identity in adolescence. In J. Adelson (Ed.) Handbook of adolescent psychology (pp. 159-189). New York: Wiley.

[46] Brickhouse, N.W. (2000). Embodying science: A feminist perspective on learning. Journal of Research in Science Teaching, 38, 282–295.

[47] Chryssochoou, X. (2003). Studying identity in social psychology: Some thoughts on the definition of identity and its relation to action. *Journal of language and Politics*, *2*(2), 225-241.

[48] Shanahan, M. C. (2009). Identity in science learning: Exploring the attention given to agency and structure in studies of identity. Studies in Science Education, 45(1), 43–64.

[49] Esteban-Guitart, M., & Moll, L. C. (2014). Funds of identity: A new concept based on the funds of knowledge approach. Culture & Psychology, 20(1), 31-48.

[50] Nasir, N. I. (2011). Racialized identities: Race and achievement among African American youth. Stanford University Press.

[51] Wenger, E. (1999). Communities of practice: Learning, meaning, and identity. Cambridge university press.

[52] Carlone, H. B., & Johnson, A. (2007). Understanding the science experiences of successful women of color: Science identity as an analytic lens. Journal of Research in Science Teaching: The Official Journal of the National Association for Research in Science Teaching, 44(8), 1187-1218.

[53] Brandt, C. B. (2008). Discursive geographies in science: Space, identity, and scientific discourse among indigenous women in higher education. Cultural Studies of Science Education, 3(3), 703-730.

[54] Rahm, J. (2007). Urban youths' identity projects and figured worlds: Case studies of youths' hybridization in science practices at the margin. Chicago, IL: The Chicago Springer Forum, Science Education in an Age of Globalization.

[55] Tonso, K. L. (2006). Student engineers and engineer identity: Campus engineer identities as figured world. Cultural studies of science education, 1(2), 273-307.

[56] Hazari, Z., Potvin, G., Lock, R. M., Lung, F., Sonnert, G., & Sadler, P. M. (2013). Factors that affect the physical science career interest of female students: Testing five common hypotheses. *Physical Review Special Topics-Physics Education Research*, *9*(2), 020115.

[57] Redish, E. F. (2010, October). Introducing students to the culture of physics: Explicating elements of the hidden curriculum. In *AIP Conference Proceedings* (Vol. 1289, No. 1, pp. 49-52). AIP.

[58] "Physics Degrees by Race/Ethnicity." American Physical Society.
https://www.aps.org/programs/education/statistics/degreesbyrace.cfm

[59] Porter, A. M. and R. Ivie. (2019). Women in Physics and Astronomy, 2019. American Institute of Physics.

[60] Wenger, E., McDermott, R. A., & Snyder, W. (2002). Cultivating communities of practice: A guide to managing knowledge. Harvard Business Press.

[61] Clandinin, D. J., & Connelly, F. M. (2000). Narrative inquiry: Experience and story in qualitative research. Wiley Press.

[62] Creswell, J. W. (2014). Creswell, Qualitative Inquiry and Research Design.